\documentclass[prd,showpacs,twocolumn,superscriptaddress,nofootinbib,floatfix,10pt]{revtex4-2}
\usepackage{setspace}
\usepackage[utf8]{inputenc}
\usepackage{makecell}
\usepackage{float}
\usepackage{amsmath,amssymb,amsfonts,bm}
\usepackage{graphicx}
\usepackage{multirow}
\usepackage{xcolor}
\definecolor{poscolor} {RGB} {252,188,190} 
\definecolor{negcolor} {RGB} {168,168,234} 
\usepackage{slashed}
\usepackage[normalem]{ulem}

\usepackage{amsmath}
\usepackage{txfonts}
\usepackage{amssymb}
\usepackage{slashed}
\usepackage{color}
\usepackage{multirow}
\usepackage{float}
\usepackage{graphicx,color,dcolumn,bm}
\usepackage{subfigure}

\usepackage{hyperref}
\hypersetup{colorlinks,citecolor=blue,anchorcolor=red,menucolor=red, linkcolor=red,filecolor=red,runcolor=red,urlcolor=blue,frenchlinks=true}
\usepackage{appendix}

\allowdisplaybreaks

\newcommand{\itp}{\affiliation{CAS Key Laboratory of Theoretical Physics, Institute of Theoretical Physics,\\ Chinese Academy of Sciences, Beijing 100190, China}}

\newcommand{\ucas}{\affiliation{School of Physical Sciences, University of Chinese Academy of Sciences, Beijing 100049, China}}

\newcommand{\imp}{\affiliation{Southern Center for Nuclear-Science Theory (SCNT), Institute of Modern Physics,\\ Chinese Academy of Sciences, Huizhou 516000, China}}

\newcommand{\xju}{\affiliation{School of Physics Science and Technology, Xinjiang University, Urumqi, Xinjiang 830046 China}}

\begin{document}
\title{Molecular states in $D_s^{(*)+}\Xi_c^{(',*)}$ systems 
}
\author{Nijiati Yalikun}
\email{nijiati@xju.edu.cn}
\xju
\author{Xiang-Kun Dong}
\email{dongxiangkun@itp.ac.cn}
\itp \ucas
\author{Bing-Song~Zou}\email{zoubs@itp.ac.cn}\itp \ucas \imp

\begin{abstract}
The possible hadronic molecules in the 
$D_s^{(*)+}\Xi_c^{(',*)}$ systems with $J^P=1/2^-,3/2^-$ and $5/2^-$ are investigated with interactions described by light meson exchanges. By varying the cutoff in a phenomenologically reasonable range of $1\sim2.5$ GeV, we find ten near-threshold (bound or virtual) states in the single-channel case. After introducing the coupled-channel dynamics of $D_s^{+}\Xi_c$-$D_s^{+}\Xi_c^{'}$-$D_s^{*+}\Xi_c$-$D_s^{+}\Xi_c^{*}$-$D_s^{*+}\Xi_c^{'}$-$D_s^{*+}\Xi_c^{*}$ systems, these states, except those below the lowest channels in each $J^{P}$ sector, move into the complex energy plane and become resonances in the mass range of $4.43\sim4.76$ GeV. Their spin-parities and nearby thresholds are $1/2^-(D_s^{+}\Xi_c)$, $1/2^-(D_s^{+}\Xi_c^{'})$, $1/2^-(D_s^{*+}\Xi_c)$,$1/2^-(D_s^{*+}\Xi_c^{'})$, $1/2^-(D_s^{*+}\Xi_c^{*})$ , $3/2^-(D_s^{*+}\Xi_c)$, $3/2^-(D_s^{+}\Xi_c)$, $3/2^-(D_s^{*+}\Xi_c^{'})$, $3/2^-(D_s^{*+}\Xi_c^{*})$, and $5/2^-(D_s^{*+}\Xi_c^{*})$. The impacts of the $\delta(\bm r)$-term in the one-boson-exchange model on these states are presented. Setting $\Lambda=1.5$ GeV as an illustrative value, it is found that $1/2^-(D_s^{+}\Xi_c)$ is a stable bound state (becoming unstable if turning on the coupling to lower channels), $1/2^-(D_s^{*+}\Xi_c)$ and $3/2^-(D_s^{*+}\Xi_c)$ are physical resonances in both cases of including or excluding the $\delta(\bm r)$-term, while the other seven states are physical resonances or ``virtual-state-like" poles near thresholds, depending on including the $\delta(\bm r)$-term or not. In addition, the partial decay widths of the physical resonances are provided. These double-charm hidden-strangeness pentaquark states, as the partners of experimentally observed $P_c$ and $P_{cs}$ states, can be searched for in the $D^{(*)}\Lambda_c$ final states in the future. 
\end{abstract}

\maketitle

\newpage 

\section{Introduction}~\label{sec:1}
The study of multiquark states began even before the birth of quantum chromodynamics (QCD), and was accelerated with the development of QCD. It is speculated that, apart from well-known $qqq$-baryons and $q\bar q$-mesons~\cite{GellMann:1964nj,Zweig:1964jf}, there would be multiquark states, glueballs, quark-gluon hybrids in the quark model notation, which are collectively called exotic hadrons.
Multiquark states can be categorized into tetraquark states ($qq\bar q \bar q$), pentaquark states ($qqq q\bar q$)
and so on. The study of multiquark states, especially how the quarks are grouped inside (i.e., compact or molecular configuration) plays a crucial role for understanding the low energy QCD.

In the past two decades, many candidates of exotic tetraquark and pentaquark states have been observed in experiments, see Refs.~\cite{Chen:2016qju,Hosaka:2016pey,Richard:2016eis,Lebed:2016hpi,Esposito:2016noz,Guo:2017jvc,Ali:2017jda,Olsen:2017bmm,Belle-II:2018jsg,Cerri:2018ypt,Liu:2019zoy,Brambilla:2019esw,Guo:2019twa,Yang:2020atz,Dong:2021juy,Chen:2022asf,Dong:2021bvy,Yamaguchi:2019vea} for recent reviews on the experimental and theoretical status of exotic hadrons. A great intriguing fact is that most of them are located quite close to the thresholds of a pair of hadrons that they can couple to. This property can be understood as there is an $S$-wave attraction between the relevant hadron pair~\cite{Dong:2020hxe}, and it naturally leads to the hadronic molecule interpretation of them (as reviewed in Refs.~\cite{Chen:2016qju,Guo:2017jvc,Brambilla:2019esw,Yamaguchi:2019vea,Dong:2021juy,Dong:2021bvy}). In the hadronic molecular picture, some of them can be interpreted as loosely bound states of two hadrons via strong interaction parameterized by light meson exchange at low energy. The validity of the hadronic molecular picture is also reflected by the successful quantitative predictions of some exotic states in early theoretical works based on the hadron-hadron  interactions, see, e.g., Refs.~\cite{Tornqvist:1993ng,Wu:2010jy,Wu:2010vk,Wang:2011rga,Yang:2011wz,Wu:2012md,Xiao:2013yca,Uchino:2015uha,Karliner:2015ina}.

The pentaquark states, $P_c(4450)$ and $P_c(4380)$, were observed by LHCb collaboration~\cite{LHCb:2015yax} in 2015. In the updated measurement~\cite{LHCb:2019kea}, the $P_c(4450)$ signal splits into two narrower peaks, $P_c(4440)$ and $P_c(4457)$ but there is no clear evidence for the previous broad $P_c(4380)$\footnote{In Ref.~\cite{Roca:2016tdh}, the results of fitting the $J/\psi p$ invariant mass distribution from the $\Lambda_b$ decay suggest the former experimental date in Ref.~\cite{LHCb:2015yax} is not enough to claim the existence of the $P_c(4380)$. }. Meanwhile, a new narrow resonance $P_c(4312)$ shows up.  Several models have been applied by tremendous works to understand the structures of these states, and the $\bar D^{(*)} \Sigma_c^{(*)}$ molecular explanation, which appeared in Refs.~\cite{Wu:2010jy,Wu:2010vk,Wu:2010rv,Wang:2011rga,Yang:2011wz,Wu:2012md,Xiao:2013yca,Karliner:2015ina} even before LHCb observations, stands out as it can explain the three states simultaneously, see, e.g., Refs.~\cite{Liu:2019tjn,Xiao:2019aya,Du:2019pij,Du:2021fmf}. The success of the hadronic molecule picture for the $P_c$ states stimulated the extension of the $\bar D^{(*)} \Sigma_c^{(*)}$ systems to their SU(3) flavor partners with hidden-charm (double-)strangeness channels~\cite{Hofmann:2005sw,Chen:2016ryt,Anisovich:2015zqa,Wang:2015wsa,Feijoo:2015kts,Lu:2016roh,Xiao:2019gjd,Chen:2015sxa,Wang:2019nvm,Zhang:2020cdi}. Recently, two $P_{cs}$ states were reported by LHCb collaboration, $P_{cs}(4459)$~\cite{LHCb:2020jpq} and $P_{cs}(4338)$~\cite{LHCb:2022jad}, which are perfect candidates of $\bar D^*\Xi_c$ and $\bar D\Xi_c$ molecules, respectively, see, e.g., Refs.~\cite{Dong:2021juy,Karliner:2022erb,Wang:2022mxy,Yan:2022wuz,Meng:2022wgl,Yang:2022ezl,Liu:2020hcv,Chen:2020kco,Wang:2020eep,Peng:2020hql,Chen:2020uif,Du:2021bgb,Xiao:2021rgp,Feijoo:2022rxf,Nakamura:2022jpd,Yan:2022wuz}.

Last year, the LHCb Collaboration announced the discovery of a double-charm exotic state, $T_{cc}^+(3985)$, which reveals itself as a high-significance peaking structure in the $D^0D^0\pi^+$ invariant mass distribution just below the nominal $D^{*+}D^0$ threshold \cite{LHCb:2021vvq,LHCb:2021auc}. This observation stimulated lots of studies of double-charm tetraquark states and $T_{cc}^+(3985)$ is a perfect candidate of the isoscalar $1^+$ $DD^*$ molecule, see, e.g., Refs.~\cite{Li:2021zbw,Du:2021zzh,Baru:2021ldu,Albaladejo:2021vln,Feijoo:2021ppq,Wang:2022jop,Ortega:2022efc,Lyu:2023xro,Chen:2022vpo,Padmanath:2022cvl}.

It is well known that the interaction between a pair of hadrons can be well described by light meson (pseudoscalar and vector) exchange. The resonance saturation has been known to be able to well approximate the low-energy constants (LECs) in the higher order Lagrangians of chiral perturbation theory~\cite{Ecker:1988te,Donoghue:1988ed}, and it turns out therein that whenever vector mesons contribute they dominate the numerical values of the LECs at the scale around the $\rho$-meson mass, which is called the modern version of vector meson dominance. Under such vector meson dominance assumption, it can be easily verified that the $D^{(*)}\Sigma_{c}^{(*)}$ systems are more attractive than the corresponding $\bar D^{(*)}\Sigma_{c}^{(*)}$ systems~\cite{Dong:2021bvy}, the latter of which correspond to the experimentally observed $P_c$ states. Such observation leads to the predictions of more deeply bound double-charm pentaquarks in the molecular scenario~\cite{Azizi:2021pbh,Dong:2021bvy,Dias:2018qhp,Shimizu:2017xrg,Chen:2021htr,Liu:2020nil,Chen:2021kad}.

In this work, we extend the study of double-charm pentaquarks to the systems with hidden-strangeness. To be specific, we explore the light-meson-exchange ($\phi,\sigma,\eta$) interactions in $D_s^{+}\Xi_c$-$D_s^{+}\Xi_c^{'}$-$D_s^{*+}\Xi_c$-$D_s^{+}\Xi_c^{*}$-$D_s^{*+}\Xi_c^{'}$-$D_s^{*+}\Xi_c^{*}$ systems and search for possible poles near the corresponding thresholds. In Sect.~\ref{sec:2}, we introduce our theoretical framework, including the involved channels, relevant Lagrangian satisfying heavy quark spin symmetry (HQSS) and SU(3) flavor symmetry, and the light-meson-exchange potentials in terms of known parameters. In Sect.~\ref{sec:res}, we show the numerical results and give some discussions. At last we give a brief summary in Sect.~\ref{sec:summary}.

\section{Theoretical framework}\label{sec:2}
The one-boson-exchange (OBE) potential model is quite successful in interpreting the formation mechanisms of pentaquarks~\cite{He:2019rva,Chen:2019asm,Liu:2019zvb,Du:2021fmf,Yalikun:2021bfm}. In this work, we also use the OBE potentials of $D_s^{(*)+}\Xi_c$, $D_s^{(*)+}\Xi_c^{'}$ and $D_s^{(*)+}\Xi_c^{*}$ systems to investigate the possibility of the double-charm pentaquarks with hidden-strangeness in the molecular picture.

\subsection{Investigated channels}\label{sec:partial-waves}
In our analysis, we focus on the hadronic molecules with spin-parities $J^P=1/2^-,3/2^-$ and $5/2^-$ in $D_s^{(*)+}\Xi_c$, $D_s^{(*)+}\Xi_c^{'}$ and $D_s^{(*)+}\Xi_c^{*}$ systems, since the negative-parity states for these channels can be coupled in $S$-wave, which is usually the most important partial wave component in a hadronic molecule. The thresholds and the spin-orbital wave functions of these five channels are listed in Table~\ref{part_waves} where the notation $^{2S+1}L_J$ is used to identify various partial waves. $S$, $L$ and $J$ stand for the total spin, orbital and total angular momenta,  respectively. The state in partial wave $^{2S+1}L_J$ with a certain $z$-direction projection $m$ can be explicitly written as
\begin{eqnarray}\label{def_state}
|LSJm\rangle=\sum\limits_{m_lm_s}\mathbb{C}_{Lm_l,Sm_s}^{Jm}|Lm_l\rangle|Sm_s\rangle,
\end{eqnarray}
where $\mathbb{C}_{Lm_l,Sm_s}^{Jm}$ is the Clebsch-Gordan coefficient, $|Sm_s\rangle$ is the spin state and $|Lm_l\rangle$ is the spatial state. In the following, we will first investigate the $S$-wave configurations to search for possible near-threshold states. Then we turn on all possible $S$-$D$-wave mixing to introduce possible $D$-wave components in each system. Other higher partial wave components, i.e., $G$-wave, are ignored due to the strong suppression from the repulsive centrifugal potential. 
\begin{table*}[htbp]\centering 
\caption{Thresholds and spin-orbital wave function of the spin-parity states $J^P$ for the $D_s^{(*)+}\Xi_c$, $D_s^{(*)+}\Xi_c^{'}$ and $D_s^{(*)+}\Xi_c^{*}$ channels. Masses of related hadrons are taken from Ref.~\cite{Zyla:2020zbs}, $m_{D_s^+}=1968.34$ MeV, $m_{D_s^{*+}}=2112.20$ MeV, $m_{\Xi_c}=2469.42$ MeV, $m_{\Xi_c'}=2578.80$ MeV and $m_{\Xi_c^*}=2645.97$ MeV.}\label{part_waves}
\begin{ruledtabular}
\begin{tabular}{c|cccccc}
Channels&$D_s^+\Xi_c$&$D_s^{+}\Xi_c^{'}$&$D_s^{*+}\Xi_c$&$D_s^+\Xi_c^*$&$D_s^{*+}\Xi_c^{'}$&$D_s^{*+}\Xi_c^*$\\\hline
Threshold [MeV] &$4437.76$&$4547.14$&$4581.62$&$4614.31$&$4691.00$&$4758.17$\\
$J^P=1/2^-$&$|^2S_{1/2}\rangle $&$|^2S_{1/2}\rangle$&$\begin{pmatrix}|^2S_{1/2}\rangle\\|^4D_{1/2}\rangle\end{pmatrix}$ &$|^4D_{1/2}\rangle$&$\begin{pmatrix}|^2S_{1/2}\rangle\\|^4D_{1/2}\rangle\end{pmatrix}$ &$\begin{pmatrix}|^2S_{1/2}\rangle\\|^4D_{1/2}\rangle\\|^6D_{1/2}\rangle\end{pmatrix}$\\
$J^P=3/2^-$&$|^2D_{3/2}\rangle $&$|^2D_{3/2}\rangle$&$\begin{pmatrix}|^4S_{3/2}\rangle\\|^2D_{3/2}\rangle\\|^4D_{3/2}\rangle\end{pmatrix}$&$\begin{pmatrix}|^4S_{3/2}\rangle\\|^4D_{3/2}\rangle\end{pmatrix}$&$\begin{pmatrix}|^4S_{3/2}\rangle\\|^2D_{3/2}\rangle\\|^4D_{3/2}\rangle\end{pmatrix}$&$\begin{pmatrix}|^4S_{3/2}\rangle\\|^2D_{3/2}\rangle\\|^4D_{1/2}\rangle\\|^6D_{3/2}\rangle\end{pmatrix}$\\
$J^P=5/2^-$&$|^2D_{5/2}\rangle $&$|^2D_{5/2}\rangle$&$\begin{pmatrix}|^2D_{5/2}\rangle\\|^4D_{5/2}\rangle\end{pmatrix}$& $|^4D_{5/2}\rangle$&$\begin{pmatrix}|^2D_{5/2}\rangle\\|^4D_{5/2}\rangle\end{pmatrix}$&$\begin{pmatrix}|^6S_{5/2}\rangle\\|^2D_{5/2}\rangle\\|^4D_{5/2}\rangle\\|^6D_{5/2}\rangle\end{pmatrix}$
\end{tabular}
\end{ruledtabular}
\end{table*}

\subsection{Effective Lagrangian and potentials}\label{subsec_Lag}

To investigate the coupling between a charmed baryon or meson with light scalar, pseudoscalar and vector mesons, we employ the effective Lagrangian satisfying chiral symmetry and HQSS, developed in Refs.~\cite{Cheng:1992xi,Yan:1992gz,Wise:1992hn,Cho:1994vg,Casalbuoni:1996pg,Pirjol:1997nh,Liu:2011xc},
\begin{align}\label{lag}
{\mathcal L}&=
l_S\bar{S}_{ab,\mu} \sigma S^\mu_{ba}
-\frac{3}{2} g_1\varepsilon_{\mu\nu\lambda\kappa}v^\kappa \bar{S}_{ab}^\mu
A_{bc}^\nu S_{ca}^\lambda\notag
\\
&+i\beta_S\bar{S}_{ab,\mu} v_\alpha (\Gamma^\alpha_{bc}-\rho^\alpha_{bc}) S^\mu_{ca}
+ \lambda_S\bar{S}_{ab,\mu} F^{\mu\nu}_{bc}S_{ca,\nu}\notag
\\
&+i\beta_B\bar B_{\bar 3Q,ab} v_\mu(\Gamma^\mu_{bc}-\rho^\mu_{bc})B_{\bar 3Q,ca}+l_B\bar B_{\bar 3Q,ab}\sigma B_{\bar{3}Q,ba}\notag\\
&+\left \{ ig_4\bar S^\mu_{ab} A_{bc,\mu} B_{\bar 3Q,ca} +i\lambda_I\varepsilon_{\mu\nu\lambda\kappa}v^\mu\bar S_{ab}^\nu F^{\lambda\kappa}_{bc}B_{\bar 3Q,ca}+h.c. \right \} \notag\\
&+i\beta{\rm Tr}[H_a^{ Q} v_\mu(\Gamma^\mu_{ab}-\rho^\mu_{ab})\bar H_b^{ Q}]+i\lambda{\rm Tr}\left[ H_a^{ Q} \frac{i}{2}[\gamma_\mu,\gamma_\nu]F^{\mu\nu}_{ab}\bar H_b^{ Q}\right]\notag\\
&+g_S{\rm Tr}[ H_a^{Q} \sigma \bar H_a^{ Q}]+ig {\rm Tr}[ H_a^{ Q} \gamma\cdot A_{ab}\gamma^5\bar H_b^{ Q}],
\end{align}
with $a,b$ and $c$ the flavor indices and $v^\mu$ the four-velocity of the heavy hadron. The $\sigma$ meson is the lightest scalar meson and is governed by the dynamics of the Goldstone bosons, relevant to the interaction between two pions~\cite{Bardeen:2003kt,Machleidt:1987hj}. The axial vector and vector currents, $A^\mu$ $\Gamma^\mu$, read
\begin{align}
A^\mu &=\frac{1}{2}(\xi^\dagger\partial^\mu\xi-\xi\partial^\mu\xi^\dagger)=\frac{i}{f_\pi}\partial^\mu\mathbb{P}+\cdots, \notag\\
\Gamma^\mu &=\frac{i}{2}(\xi^\dagger\partial^\mu\xi+\xi\partial^\mu\xi^\dagger)=\frac{i}{2f_\pi^2}[\mathbb{P},\partial^\mu\mathbb{P}]+\cdots, 
\end{align}
with $\xi={\rm exp}(i\mathbb{P}/f_\pi)$. $f_\pi=132$~MeV is the pion decay constant. The vector meson fields $\rho^\alpha$ and field strength tensor $F^{\alpha\beta}$ are defined as $\rho^\alpha={ig_V}\mathbb{V}^\alpha/{\sqrt 2}$ and $F^{\alpha\beta}=\partial^\alpha\rho^\beta-\partial^\beta\rho^\alpha+[\rho^\alpha,\rho^\beta]$. $\mathbb{P}$ and $\mathbb{V}^\alpha$ denote the light pseudoscalar octet and the light vector nonet, respectively, 
\begin{align}
\mathbb{P}&=
\begin{pmatrix}
\frac{\pi^0}{\sqrt 2}+\frac{\eta}{\sqrt 6}&\pi^+&K^+\\
\pi^-&-\frac{\pi^0}{\sqrt 2}+\frac{\eta}{\sqrt 6}&K^0\\
K^-&\bar K^0&-\sqrt{\frac{2}{3}}\eta
\end{pmatrix},\\
\mathbb{V}&=
\begin{pmatrix}
\frac{\rho^0}{\sqrt 2}+\frac{\omega}{\sqrt 2}&\rho^+&K^{*+}\\
\rho^-&-\frac{\rho^0}{\sqrt 2}+\frac{\omega}{\sqrt 2}&K^{*0}\\
K^{*-}&\bar K^{*0}&\phi
\end{pmatrix},
\end{align}
where we have ignored the mixing between pseudoscalar octet and singlet. The $S$-wave heavy meson $Q\bar q$ and baryon $Qqq$ containing a single heavy quark can be represented with interpolated fields $H_a^{Q}$ and $S_{ab}^\mu$, respectively. 
\begin{align}
H_a^{Q}&=\frac{1+\slashed v}{2}(\mathcal{P}^*_{a,\mu}\gamma^\mu-\mathcal{P}_a\gamma^5),\\
\bar H_a^{ Q}&=\gamma^0H_a^{ Q\dagger}\gamma^0,\\
S^\mu_{ab}&=-\frac{1}{\sqrt3}(\gamma^\mu+v^\mu)\gamma^5 (B_{6Q})_{ab}+(B_{6Q}^{*,\mu})_{ab},\\
\bar S^\mu_{ab}&=S^{\mu\dagger}_{ab}\gamma^0,
\end{align}
where heavy mesons with $J^P=0^-$ and $J^P=1^-$ are denoted by $\mathcal{P}$ and $\mathcal{P}^*_\mu$, respectively, while the heavy baryons with $J^P=1/2^+$ and $3/2^+$ in the $6_{\rm{F}}$ representation of SU(3) for the light quark flavor symmetry are labelled by $B_{6Q}$ and $B_{6Q}^{*,\mu}$. For the case of $Q=c$, they are written in the SU(3) flavor multiplets as 
\begin{align}
\mathcal{P}&=(D^0,D^+,D_s^+), \qquad \mathcal{P}^*=(D^{*0},D^{*+},D_s^{*+}),\\
B_{6c}&=
\begin{pmatrix}
\Sigma_c^{++}&\Sigma_c^{+}/\sqrt 2&\Xi_c^{'+}/\sqrt 2\\
\Sigma_c^{+}/\sqrt 2&\Sigma_c^{0}&\Xi_c^{'0}/\sqrt 2\\
\Xi_c^{'+}/\sqrt 2&\Xi_c^{'0}/\sqrt 2&\Omega^{0}_c
\end{pmatrix},\\
B_{6c}^{*}&=
\begin{pmatrix}
\Sigma_c^{*++}&\Sigma_c^{*+}/\sqrt 2&\Xi_c^{*+}/\sqrt 2\\
\Sigma_c^{*+}/\sqrt 2&\Sigma_c^{*0}&\Xi_c^{*0}/\sqrt 2\\
\Xi_c^{*+}/\sqrt 2&\Xi_c^{*0}/\sqrt 2&\Omega^{*0}_c
\end{pmatrix}.
\end{align}
The $S$-wave heavy baryons with $J^P=1/2^+$ in $\bar 3_{\rm{F}}$ representation are embedded in
\begin{align}
B_{\bar 3c}&=\left(\begin{array}{ccc}
0&\Lambda_{c}^+&\Xi^+_{c}\\
-\Lambda_{c}^+&0&\Xi^0_{c}\\
-\Xi^+_{c}&-\Xi^0_{c}&0
\end{array}\right).
\end{align}

With the Lagrangian in Eq.~\eqref{lag}, we can derive the analytic expressions of the potentials describing the OBE dynamics for the  $D_s^{+}\Xi_c$, $D_s^{+}\Xi_c^{'}$, $D_s^{*+}\Xi_c$, $D_s^{+}\Xi_c^{*}$, $D_s^{*+}\Xi_c^{'}$ and $D_s^{*+}\Xi_c^{*}$ systems. By Breit approximation,the potential in momentum space reads
\begin{align}
\mathcal{V}^{h_1h_2\to h_3h_4}(\bm q)=-\frac{\mathcal{M}^{h_1h_2\to h_3h_4}}{\sqrt{2m_12m_22m_32m_4}},\label{eq:Breit}
\end{align}
 where $m_i$ is the mass of the particle $h_i$, $\bm q$ is the three momentum of the exchanged meson and
 $\mathcal{V}^{h_1h_2\to h_3h_4}$ is the scattering amplitude of the transition $h_1h_2\to h_3 h_4$. In our calculation, spinors of spin-$1/2$ and $3/2$ fermions with positive energy in nonrelativistic approximation read~\cite{Lu:2017dvm}, 
 \begin{align}
u(p,m)_{B_{\bar 3 c}/B_{6 c}}&=\sqrt{2M_{B_{\bar 3 c}/B_{6 c}}}\begin{pmatrix}
\chi_m\\0
\end{pmatrix},\\
u(p,m)_{B^*_{6 c}}&=\sqrt{2M_{B^*_{6 c}}}\begin{pmatrix}
(0,\bm\chi_{m})\\(0,\bm 0)
\end{pmatrix},
 \end{align} 
 where $\chi_m$ is the two-component spinor and
\begin{align}
\bm{\chi}_m=\sum_{m_1,m_2}\mathbb{C}_{1,m_1;1/2,m_2}^{3/2,m}\bm\epsilon(m_1)\chi_{m_2},
\end{align} 
with $\bm\epsilon(\pm 1)=(\mp 1,-i,0)/\sqrt{2}$ and $\bm\epsilon(0)=(0,0,1)$. The scaled heavy meson fields $\mathcal{P}$ and $\mathcal{P}^*$ are normalized as~\cite{Wise:1992hn,Wang:2020dya}
\begin{align}
    \langle 0 |\mathcal{P}| c \bar q(0^-)\rangle=\sqrt{M_{\mathcal{P}}},\quad    \langle 0|\mathcal{P}^*_\mu|c \bar q(1^-)\rangle=\epsilon_\mu\sqrt{M_{\mathcal{P}^*}}. 
\end{align} 
For convenience, the six channels, $D_s^{+}\Xi_c$, $D_s^{+}\Xi_c^{'}$, $D_s^{*+}\Xi_c$, $D_s^{+}\Xi_c^{*}$, $D_s^{*+}\Xi_c^{'}$ and $D_s^{*+}\Xi_c^{*}$ are labeled as channel $1\sim 6$, respectively, sorted by their thresholds. The OBE potentials in the momentum space, $\mathcal{V}^{ij}$ for $i\to j$ channel transition, are derived in the center of mass frame and shown explicitly in Appendix~\ref{sec:potential}.

The potentials in coordinate space are obtained by performing the Fourier transformation,
\begin{align}
\mathcal{V}(\bm r,\Lambda,\mu_{\rm{ex}})=\int\frac{d^3\bm q}{(2\pi)^3} \mathcal{V}(\bm q)F^2(\bm q,\Lambda,\mu_{\rm{ex}}){\rm{e}}^{i\bm q \cdot \bm r},\label{eq:Fourier-trans}
\end{align} 
where the form factor with the cutoff $\Lambda$ is introduced to account for the inner structures of the interacting hadrons~\cite{Tornqvist:1993ng},
\begin{align}
F(\bm q,\Lambda,\mu_{\rm{ex}})=\frac{m_{\rm{ex}}^2-\Lambda^2}{(q^0)^2-\bm q^2-\Lambda^2}=\frac{\tilde{\Lambda}^2-\mu_{\rm{ex}}^2}{\bm q^2+\tilde{\Lambda}^2}.\label{eq:form-factor}
\end{align}
We have defined $\tilde\Lambda=\sqrt{\Lambda^2-(q^0)^2}$ and $\mu_{\rm{ex}}=\sqrt{m_{\rm{ex}}^2-(q^0)^2}$ for convenience. Note that, for the inelastic scattering, the energy of the exchanged meson is nonzero, so the denominator of the propagator can be rewritten as $q^2-m_{\rm{ex}}^2=(q^0)^2-\bm q^2 - m_{\rm{ex}}^2=-(\bm q^2+\mu_{\rm{ex}}^2)$, with $\mu_{\rm{ex}}$ the effective mass of the exchanged meson. The energy of the exchanged meson $q^0$ is calculated nonrelativistically as
\begin{align}
q^0=\frac{m_2^2-m_1^2+m_3^2-m_4^2}{2(m_3+m_4)},
\end{align}  
where $m_1(m_3)$ and $m_2(m_4)$ are the masses of the charmed-baryon and -meson in the initial(final) state. The momentum space potentials in Eqs.~\eqref{eq:poten-in-p}  in Appendix~\ref{sec:potential} include three types of functions, $1/(\bm q^2+\mu^2_{\rm {ex}})$, $\bm A\cdot \bm q \bm B\cdot \bm q/(\bm q^2+\mu^2_{\rm {ex}})$ and $(\bm A\times \bm q) \cdot(\bm B\times \bm q)/(\bm q^2+\mu^2_{\rm {ex}})$. $\bm A$ and $\bm B$ refer to the vector operators acting on the spin-orbit wave functions of the initial or final states, and their specific forms can be deduced from the corresponding terms in Eqs.~\eqref{eq:poten-in-p}. For instance, $\bm A=\chi^\dagger_3 \bm\sigma\chi_1$ and $\bm B=\bm\epsilon_4^*$ in Eq.~\eqref{eq:p-v15}. The Fourier transformation of $1/(\bm q^2+\mu_{\rm {ex}})$, denoted by $Y_{\rm{ex}}$, reads
\begin{align}
Y_{\rm{ex}}&=\int\frac{d^3\bm q}{(2\pi)^3} \frac{1}{\bm q^2+\mu_{\rm{ex}}} \left (\frac{\tilde{\Lambda}^2-\mu^2_{\rm{ex}}}{\bm q^2+\tilde{\Lambda}^2}\right )^2 e^{i\bm q\cdot \bm r},\notag\\
&=\frac{1}{4\pi r}({\rm{e}}^{-\mu_{\rm{ex}}r}-{\rm{e}}^{-\tilde{\Lambda}r})-\frac{\tilde{\Lambda}^2-\mu_{\rm{ex}}^2}{8\pi\tilde{\Lambda}}{\rm{e}}^{-\tilde{\Lambda}r}.\label{eq:Yex}
\end{align}     
Before performing the Fourier transformation on $\bm A\cdot \bm q \bm B\cdot \bm q/(\bm q^2+\mu_{\rm{ex}}^2)$, we can decompose it as 
\begin{align}
\frac{\bm A\cdot \bm q \bm B\cdot \bm q}{\bm q^2+\mu_{\rm{ex}}^2}&=\frac{1}{3}\left \{\bm A\cdot\bm B\left (1-\frac{\mu_{\rm{ex}}^2}{\bm q^2+\mu_{\rm{ex}}^2}\right )-\frac{S(\bm A,\bm B,\hat q)|\bm q|^2}{\bm q^2+\mu_{\rm{ex}}^2}\right \}, \label{eq:CT-pot}
\end{align}
where $S(\bm A,\bm B,\hat q)=3 \bm A\cdot \hat q \bm B\cdot \hat q-\bm A\cdot\bm B$ is the tensor operator in momentum space. It can be found that without the form factor the constant term in Eq.~\eqref{eq:CT-pot} leads to a $\delta(\bm r)$-term in coordinate space after the Fourier transformation. With the form factor, the $\delta(\bm r)$-term becomes finite, and it dominates the short-range part of the potential. In the phenomenological view, the $\delta(\bm r)$-term can mimic the role of contact interaction~\cite{Yalikun:2021bfm}, which is also related to the regularization scheme~\cite{Tornqvist:1993ng}. In Refs.~\cite{Thomas:2008ja,Yamaguchi:2017zmn}, after removing $\delta(\bm r)$-term, the hadronic molecular picture for some observed hidden-charm states is discussed with the pion-exchange potential which is assumed to be of long-range. In this work, we will separately analyze the poles in the system with or without the effect of the $\delta(\bm r)$-term. For this propose, we introduce a parameter $a$ to distinguish these two case, 
\begin{align}
\frac{\bm A\cdot \bm q \bm B\cdot \bm q}{\bm q^2+\mu_{\rm{ex}}^2}-\frac{a}{3}\bm A\cdot \bm B&=\frac{1}{3}\left \{\bm A\cdot\bm B\left (1-a-\frac{\mu_{\rm{ex}}^2}{\bm q^2+\mu_{\rm{ex}}^2}\right )\right.\notag\\
&\ \ \ \ \left.-S(\bm A,\bm B,\hat q)\frac{|\bm q|^2}{\bm q^2+\mu_{\rm{ex}}^2}\right \}. \label{eq:CT-pot1}
\end{align}
After performing the Fourier transformation of Eq.~\eqref{eq:CT-pot1}, we have
\begin{align}
&\int\frac{d^3\bm q}{(2\pi)^3} \left (\frac{\bm A\cdot \bm q \bm B\cdot \bm q}{\bm q^2+\mu_{\rm{ex}}^2}-\frac{a}{3}\bm A\cdot \bm B\right ) \left (\frac{\tilde{\Lambda}^2-\mu^2_{\rm{ex}}}{\bm q^2+\tilde{\Lambda}^2}\right )^2 e^{i\bm q\cdot \bm r}\notag\\
&=-\frac{1}{3}[\bm A\cdot \bm BC_{\rm{ex}}+S(\bm A,\bm B,\hat r)T_{\rm{ex}}]\label{eq:CT-pot2},
\end{align}
where $S(\bm A,\bm B,\hat r)=3 \bm A\cdot \hat r \bm B\cdot \hat r-\bm A\cdot\bm B$ is the tensor operator in coordinate space, and the functions $C_{\rm{ex}}$ and $T_{\rm{ex}}$ read 
\begin{align}
C_{\rm{ex}}&=\frac{1}{r^2}\frac{\partial}{\partial r}r^2\frac{\partial}{\partial r}Y_{\rm{ex}}+\frac{a}{(2\pi)^3}\int \left (\frac{\tilde{\Lambda}^2-\mu^2_{\rm{ex}}}{\bm q^2+\tilde{\Lambda}^2}\right )^2 e^{i\bm q\cdot \bm r}d^3\bm q,\\
T_{\rm{ex}}&=r\frac{\partial}{\partial r}\frac{1}{r}\frac{\partial}{\partial r}Y_{\rm{ex}}.
 \end{align}   
Apparently, the contribution of the $\delta(\bm r)$-term is fully included (excluded) when $a=0(1)$~\cite{Yalikun:2021bfm,Wang:2020dya}.\footnote{From the perspective of effective field theory, such short-range interactions actually serve as the counter terms for the renormalization to cancel the cutoff dependence of the pole positions. It was found that the parameter $a={\cal O}(1)$ in the previous studies of $P_{c}$ states as molecules of $\bar D^{(*)}\Sigma_c^{(*)}$~\cite{Yalikun:2021bfm} and charmonium-like states as molecules of $D^{(*)}\bar D_{(1,2)}$~\cite{Ji:2022blw}.} Similarly, the Fourier transformation of the function $(\bm A\times \bm q) \cdot(\bm B\times \bm q)/(\bm q^2+\mu^2_{\rm {ex}})$ can be evaluated with the help of the relation $(\bm A\times \bm q) \cdot(\bm B\times \bm q)=\bm A\cdot \bm B|\bm q|^2-\bm A\cdot \bm q \bm B\cdot \bm q$. 

With the prescription above, the coordinate space representations of the potentials in Eqs.~\eqref{eq:poten-in-p} can be written in terms of the functions $Y_{\rm{ex}}$, $C_{\rm{ex}}$ and $T_{\rm{ex}}$ given in Eqs.~\eqref{eq:Yex} and ~\eqref{eq:CT-pot2}. The potentials should be projected into certain partial waves by sandwiching the spin operators in the potentials between the partial waves of the initial and final states. Computing the partial wave projection is travail and we refer to Refs.~\cite{Yalikun:2021bfm,Yalikun:2021dpk} for details. 

In our calculations, the masses of exchanged particles are $m_{\sigma}=600.0$ MeV, $m_{\eta}=547.9$ MeV and $m_{\phi}=1019.5$ MeV. The coupling constants in the Lagrangian can be extracted from experimental data or deduced from various theoretical models. Here we adopt the values given in Refs.~\cite{Ding:2008gr,Liu:2011xc,Meng:2019ilv,Isola:2003fh}, $l_S=6.20$, $g_S=0.76$, $l_B=-3.65$, $g=-0.59$, $g_1=0.94$, $g_4=1.06$, $\beta g_V=-5.25$, $\beta_S g_V=10.14$, $\beta_B g_V=-6.00$, $\lambda g_V=-3.27~\rm{GeV}^{-1}$, $\lambda_s g_V=19.2~\rm{GeV}^{-1}$, and $\lambda_I g_V=-6.80~\rm{GeV}^{-1}$, while their relative phases are fixed by the quark model~\cite{Riska:2000gd,Yalikun:2021dpk}, .    

\begin{figure*}[ht]\centering
\includegraphics[width=0.85\textwidth]{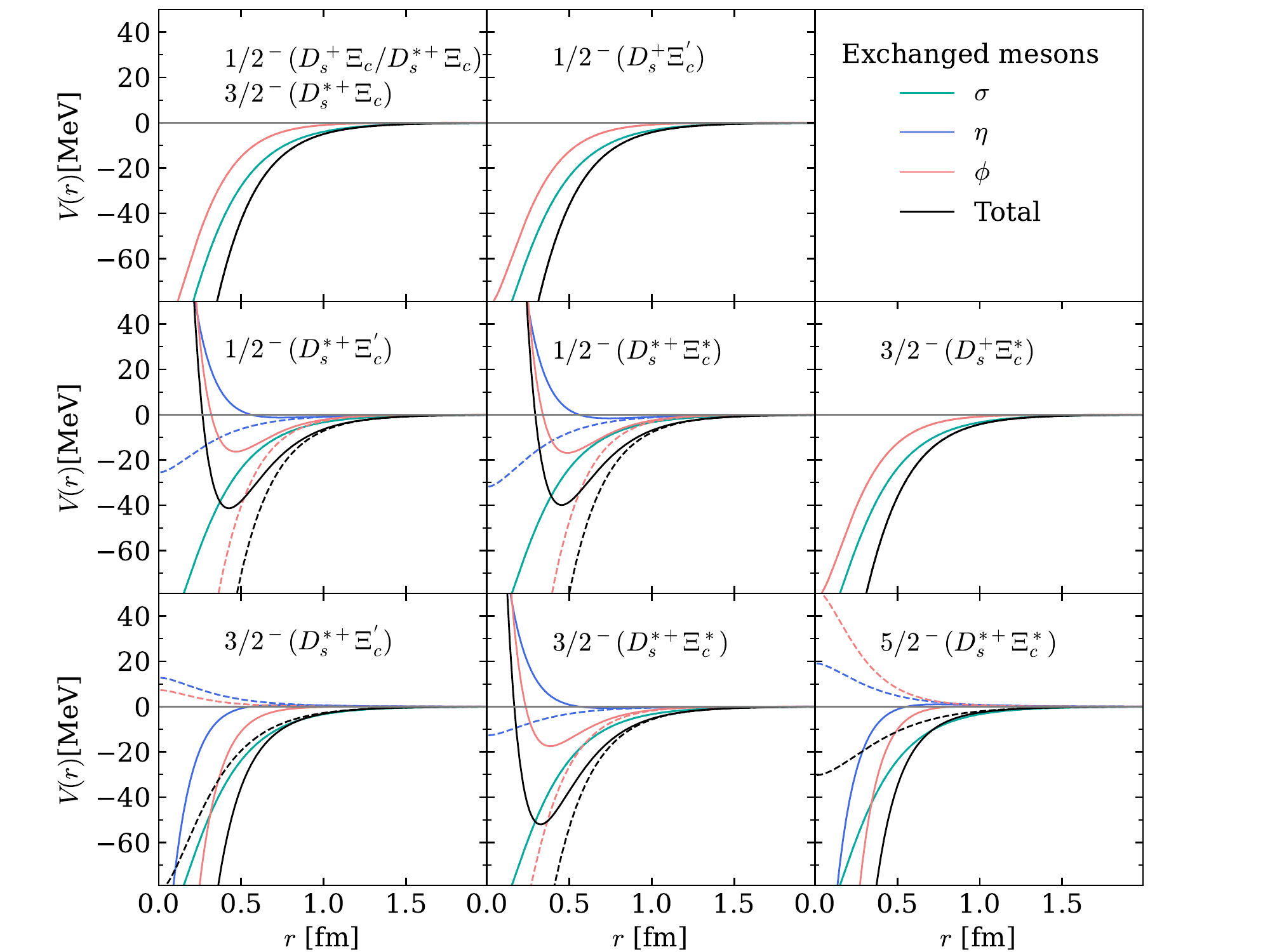}
\caption{The $S$-wave potentials in single channels with $\Lambda=1.5$ GeV. Solid(dashed) lines correspond to $a=0(1)$, i.e., with(without) the $\delta(\bm r)$-term.}\label{fig_pot}
\end{figure*}

The possible $S$-wave potentials of the six channels with $\Lambda=1.5$ GeV are shown in Fig.~\ref{fig_pot}. 
It is seen that the $S$-wave potentials of $D_s^{+}\Xi_c$, $D_s^{+}\Xi_c^{'}$, $D_s^{*+}\Xi_c$ and $D_s^{+}\Xi_c^*$ channels are only proportional to the Yukawa-type potential $Y(r,\Lambda,m_{\rm{ex}})$, and thus they are independent of the $\delta(\bm r)$-term.  Potentials of other channels apart from $\sigma$ exchange depend on the $\delta(\bm r)$-term and thus the short-range (smaller than 1 fm roughly) potentials have different shapes for $a=0$ or 1. We can also see that if the short-range potentials depend on the $\delta(\bm r)$-term, they are dominated by the $\delta(\bm r)$-term.  

\subsection{Schr\"odinger equations and poles}
The LHCb $P_c$ pentaquarks were discovered in the analysis of the
invariant mass distributions of $J/\psi p$, and their masses are several MeV below the thresholds of $\bar D^{(*)}\Sigma_c$ systems~\cite{LHCb:2015yax,LHCb:2019kea}. A natural explanation is that the pentaquarks arise as bound states of $\bar D^{(*)}\Sigma_c^{(*)}$, in which the nonrelativistic potentials for time-independent Schr\"odinger equation deduced from $t$-channel scattering amplitude is a good description of the interaction of these systems~\cite{Chen:2019asm,Liu:2019tjn,Xiao:2019aya,Du:2019pij,Du:2021fmf,Yalikun:2021bfm}. We use the nonrelativistic potentials derived in the previous subsection to explore bound states or resonances in $D^{(*)+}\Xi_c^{('*)}$ systems. For the coupled-channel potential matrix $\mathcal{V}_{jk}$, the radial Schr\"odinger equation can be written as  
\begin{eqnarray}
\left[-\frac{1}{2\mu_j}\frac{d^2}{d r^2}+\frac{l_j(l_j+1)}{2\mu_jr^2} + W_j\right]u_j+
\sum_{k}
\mathcal{V}_{jk}u_k=
E u_j,
\label{eq_schro_coupl}
\end{eqnarray}
where $j$ is the channel index; $u_j$ is defined by $u_j(r)=rR_j(r)$ with the radial wave function $R_j(r)$ for the $j$-th channel; $\mu_j$ and $W_j$ are the corresponding reduced mass and threshold; $E$ is the total energy of the system. The momentum for channel $j$ is expressed as 
\begin{align}\label{eq:ch-mom}
q_j(E)=\sqrt{2\mu_j(E-W_j)}.
\end{align}
 By solving Eq.~\eqref{eq_schro_coupl}, we obtain the wave function which is normalized to satisfy the incoming boundary condition for the $j$-th channel~\cite{osti_4661960},
\begin{eqnarray}
u_{j}^{(k)}(r)\overset{r\rightarrow \infty}{\longrightarrow} \delta_{jk}e^{-iq_j r}-S_{jk}(E)e^{iq_j r}\label{eq:asym-wave},
\end{eqnarray}
where $S_{jk}(E)$ is the scattering matrix component. In multi-channel problem, there is a sequence of thresholds, $W_1<W_2<\cdots$, and the scattering matrix element $S_{jk}(E)$ is analytic function of $E$ except at the branch points $E=W_j$ and possible poles. Bound/virtual states and resonances are represented as the poles of the $S_{jk}(E)$ in the complex energy plane~\cite{osti_4661960}. 

The characterization of these poles requires to analytically continue the $S$ matrix to the complex energy plane, and the poles should be searched on the correct Riemann sheet (RS). Note that momentum $q_j$ is a double-valued function of energy $E$ and there are two RSs in the complex energy plane for each channel, one called the first or physical sheet and the other called the second or unphysical sheet. In the physical sheet, complex energy $E$ maps to the upper-half plane (${\rm{Im}}[q_j]\geq 0$) of $q_j$. In the unphysical sheet, complex energy $E$ maps to the lower-half plane (${\rm{Im}}[q_j]< 0$) of $q_j$. In the coupled-channel system with $n$ channels, the scattering amplitude has $2^n$ RSs, which can be defined by the imaginary part of the momentum $q_j(E)$ of the $j$-th channel (see the chapter 20 of Ref.~\cite{osti_4661960} for more details). Each RS is labeled by $r=(\pm,\cdots,\pm)$ and the $j$-th ``$\pm$" here stands for the sign of the imaginary parts of the $j$-th channel momentum $q_j(E)$.

\section{Results and discussions}\label{sec:res}
\subsection{Single-channel analysis}
Now, we are ready to discuss the possibility of bound {or virtual} states in $D_s^{+}\Xi_c$, $D_s^{+}\Xi_c^{'}$, $D_s^{*+}\Xi_c$, $D_s^{+}\Xi_c^{*}$, $D_s^{*+}\Xi_c^{'}$ and $D_s^{*+}\Xi_c^{*}$ systems by varying $\Lambda$ in the range of $1.0\sim2.5$ GeV. Considering the OBE potentials and $S$-$D$-wave mixing, the pole positions are obtained by solving the the Schr\"odinger equation in Eq.~\eqref{eq_schro_coupl}. As discussed in previous section, the $\delta(\bm r)$-term dominates the short-range dynamics of the potentials, and thus it serves as the phenomenological contact term which is used to determine the short-range dynamics of hadron interactions~\cite{Du:2021fmf}. It is seen that the the proper treatment of $\delta(\bm r)$ in OBE model plays an important role in the simultaneous interpretation of the LHCb $P_c$ states~\cite{Yalikun:2021bfm}. Therefore, we will represent the results in two extreme cases with $a=0$ or 1. 

 In the single-channel case, the bound state corresponds to the pole located at the real energy axis below the threshold on the first RS, while the virtual state corresponds to the pole at the real energy axis below the threshold on the second RS. The binding energies of the bound or virtual states are defined as 
 \begin{align}
 \mathbb{B}=E_{\mathrm{pole}}-W.
 \end{align}
 
\begin{figure}[ht]\centering
\includegraphics[width=\linewidth]{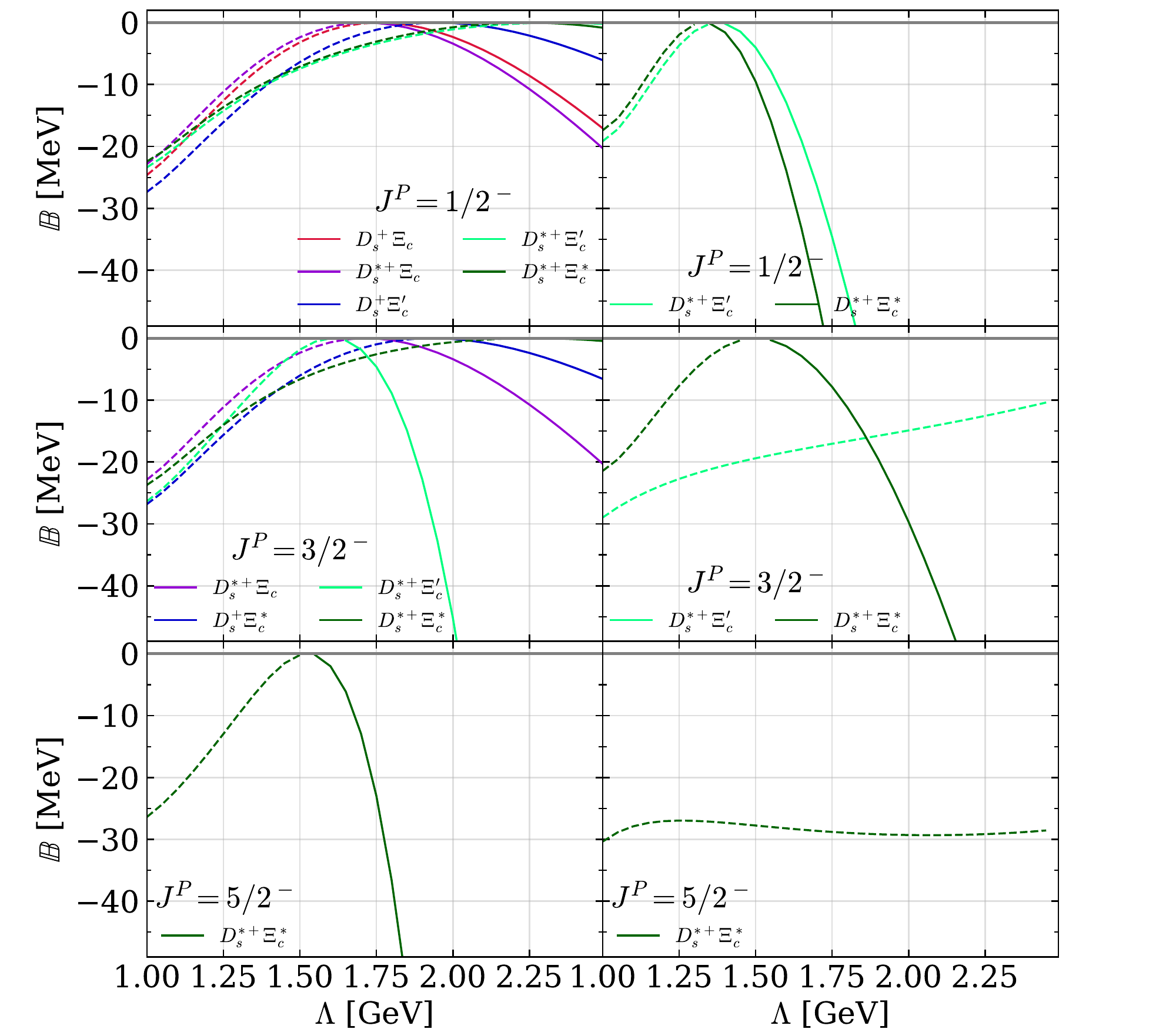}
\caption{The binding energy ($\mathbb{B}$) of the bound states (solid curves) or virtual states (dashed curves) in the single channels as $\Lambda$ increases. The results without $\delta(\bm r)$-term are shown in the left panel.}\label{fig:BS-VS}
\end{figure}

 In the single-channel case, the binding energies for these systems with $J^P=1/2^-,3/2^-$ and $5/2^-$ when the cutoff varies from $1.0$ to $2.5$ GeV are shown in Fig.~\ref{fig:BS-VS}. Complementary to that, as we have seen in Fig.~\ref{fig_pot} that the OBE potentials for  $D_s^{*+}\Xi_c^{'}$ and $D_s^{*+}\Xi_c^{*}$ channels depend on the $\delta(\bm r)$-term, while that of other channels are free of the $\delta(\bm r)$-term, the results in these two channels when the $\delta(\bm r)$-term is removed are given in the three subplots at the right panel of Fig.~\ref{fig:BS-VS}. As shown in the three subplots at the left panel of Fig.~\ref{fig:BS-VS}, ten virtual states are found when $\Lambda=1.0$ GeV, and they become bound states as the cutoff increases. In the $J^P=1/2^-$ sector, $D_s^{+}\Xi_c$, $D_s^{*+}\Xi_c$ and $D_s^{+}\Xi_c^{'}$ states are more easily bound compared to those in $D_s^{*+}\Xi_c^{'}$ and $D_s^{*+}\Xi_c^{*}$ channels because the $\delta(\bm r)$-terms in the potentials of the latter two channels are repulsive. After removing the $\delta(\bm r)$-term, both $D_s^{*+}\Xi_c^{'}$ and $D_s^{*+}\Xi_c^{*}$ channels can form relatively deep bound states, as shown in the right-top subplot in Fig.~\ref{fig:BS-VS}. However, the $J^P=1/2^-$ $D_s^{+}\Xi_c^{*}$ system is in $D$-wave and can not form a bound state. In the $J^P=3/2^-$ sector, the formation of the $D_s^{*+}\Xi_c^{'}$ and $D_s^{*+}\Xi_c^*$ bound states is sensitive to the treatment of the $\delta(\bm r)$-term. For instance, $D_s^{*+}\Xi_c^{'}$ system is more difficult to be bound when the $\delta(\bm r)$-term is removed but the situation is reversed for the $D_s^{*+}\Xi_c^*$ system, due to the opposite sign of the $\delta(\bm r)$-term in these two systems. For the $J^P=5/2^-$ system, only one near-threshold pole is found, corresponding to the $D_s^{*+}\Xi_c^*$ channel. Including the $\delta(\bm r)$-term in this channel make the binding easier.   

 It is found in Ref.~\cite{Dong:2021bvy} that when only the $\phi$ meson exchange is considered, the poles in the ten channels we mentioned above are located at the second RS below the corresponding thresholds, and they move toward thresholds as the cutoff increases in a reasonable range. In this work, we consider the contribution from other mesons exchange, $\eta$ and $\sigma$, and hence the more attractive potentials used in our work together with $S-D$ wave mixing effect naturally push these poles on the second sheets to the 1st RS when cutoff is increased. In addition, the formation of the bound states in double charm  and hidden strangeness systems, $D_s^{(*)+}\Xi_c^{(',*)}$, is easier than that in hidden charm  and double strangeness systems, as investigated in Ref.~\cite{Wang:2020bjt}.   

\subsection{Coupled-channel analysis}
We further investigate the coupled-channel dynamics of $D_s^{+}\Xi_c$-$D_s^{+}\Xi_c^{'}$-$D_s^{*+}\Xi_c$-$D_s^{+}\Xi_c^{*}$-$D_s^{*+}\Xi_c^{'}$-$D_s^{*+}\Xi_c^{*}$ system via solving the Schr\"odinger equation in Eq.~\eqref{eq_schro_coupl}. Physical resonances are calculated by analytic continuation of the $S(E)$ matrix extracted from the asymptotic wave function in Eq.~\eqref{eq:asym-wave} (see Refs.~\cite{Kamiya:2019uiw,osti_4661960}). In our case of the 6-channel system, there are $2^6$ Riemann sheets, labeled by $r=(\pm\pm\pm\pm\pm\pm)$. Note that we will only focus on several of them that are relatively close to the physical real axis. We refer to review section of Ref.~\cite{Zyla:2020zbs} for connections between each RS to the physical real axis.

Since the $S$-wave component of the coupled channels is important for near-threshold poles and contributions from other higher partial wave components are highly suppressed by the centrifugal potentials,  we first turn off the $S$-$D$-wave mixing and only consider the $S$-wave components to see the pole trajectories in the $D_s^{+}\Xi_c$-$D_s^{+}\Xi_c^{'}$-$D_s^{*+}\Xi_c$-$D_s^{+}\Xi_c^{*}$-$D_s^{*+}\Xi_c^{'}$-$D_s^{*+}\Xi_c^{*}$ coupled-channel system by varying $\Lambda$.

\begin{figure}[th]\centering
	\includegraphics[width=0.48\textwidth]{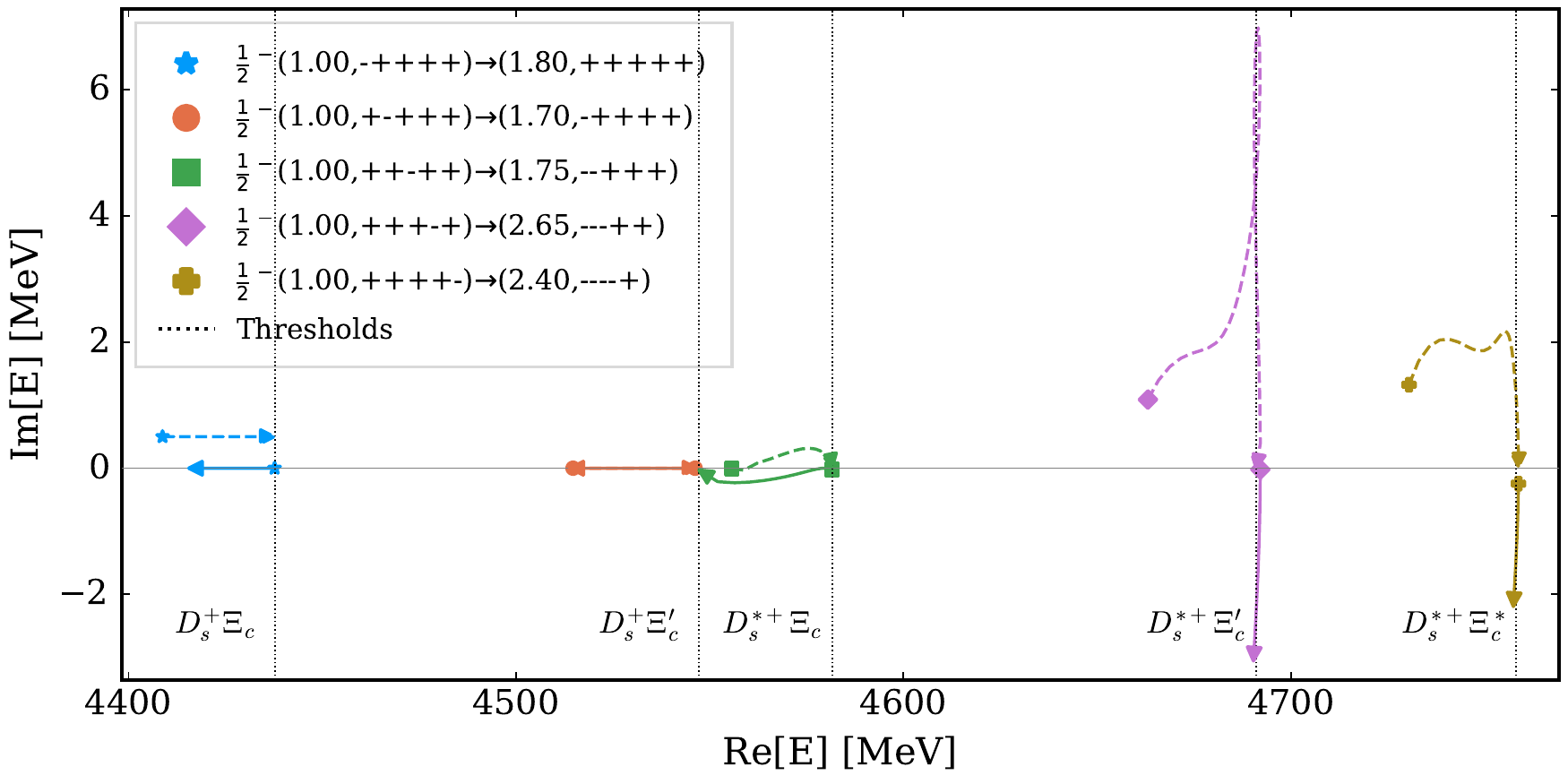}
	\caption{Trajectories of the near-threshold poles in $D_s^{+}\Xi_c$-$D_s^{+}\Xi_c^{'}$-$D_s^{*+}\Xi_c$-$D_s^{*+}\Xi_c^{'}$-$D_s^{*+}\Xi_c^{*}$ channels with $J^P=1/2^-$ by varying the cutoff. For each pole, the dashed(solid) curve represents the trajectory of the pole in the RS, whose label is shown in the left(right) parenthesis in the legend, where the number (in unit of GeV) is the starting value of the cutoff. The trajectory of the virtual pole of $D_s^{+}\Xi_c$ system is artificially moved from the real axis to the complex plane for better illustration.}\label{fig:RSS10}
\end{figure}

\begin{figure}[ht]\centering
	\includegraphics[width=0.48\textwidth]{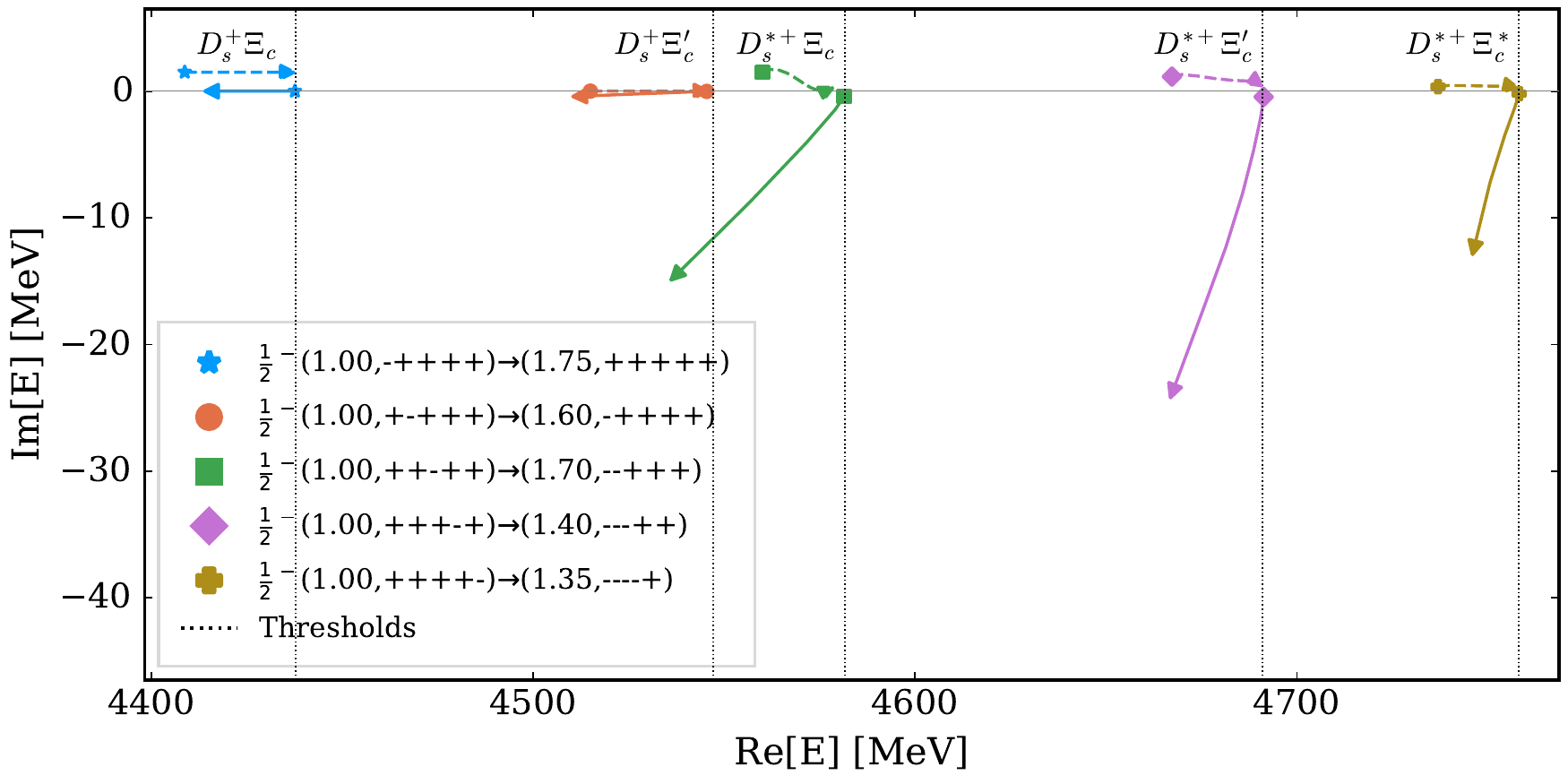}
	\caption{Similar figure as Fig.~\ref{fig:RSS10} after removing the $\delta(\bm r)$-term. }\label{fig:RSS11}
\end{figure}

For the $J^P=1/2^-$ system, five channels, $D_s^{+}\Xi_c$-$D_s^{+}\Xi_c^{'}$-$D_s^{*+}\Xi_c$-$D_s^{*+}\Xi_c^{'}$-$D_s^{*+}\Xi_c^{*}$, can couple in $S$-wave. In this case, the trajectories of poles near thresholds of these five channels as the cutoff increases from $1.0$ GeV are shown in Fig.~\ref{fig:RSS10}, and the similar results after removing the $\delta(\bm r)$-term from the potentials are shown in Fig.~\ref{fig:RSS11}. When $\Lambda=1.0$ GeV, five near-threshold poles emerge simultaneously on the complex plane, below the thresholds of the five channels, respectively, but they are not connected to the physical real axis directly. They move to the right and approach to the thresholds as the cutoff increases. If the cutoff increases up to sufficiently large values, such that $1.80,~1.70,~1.75,~2.65$ and $2.40$ GeV for the five poles below thresholds of $D_s^{+}\Xi_c$, $D_s^{+}\Xi_c^{'}$, $D_s^{*+}\Xi_c$, $D_s^{*+}\Xi_c^{'}$ and $D_s^{*+}\Xi_c^{*}$ channels, respectively, these five poles move into other RSs and get connected to the physical real axis directly. As shown in Fig.~\ref{fig:RSS11}, if the $\delta(\bm r)$-terms are removed from the potential, the poles near thresholds of $D_s^{*+}\Xi_c^{'}$ and $D_s^{*+}\Xi_c^{*}$ appear in the region connected to the physical real axis with smaller cutoff. Such behavior of these two poles from the effect of the $\delta(\bm r)$-term also mimics that of bound states in the single-channel case of $D_s^{*+}\Xi_c^{'}$ and $D_s^{*+}\Xi_c^{*}$ in the previous subsection.

\begin{figure}[ht]\centering
	\includegraphics[width=0.48\textwidth]{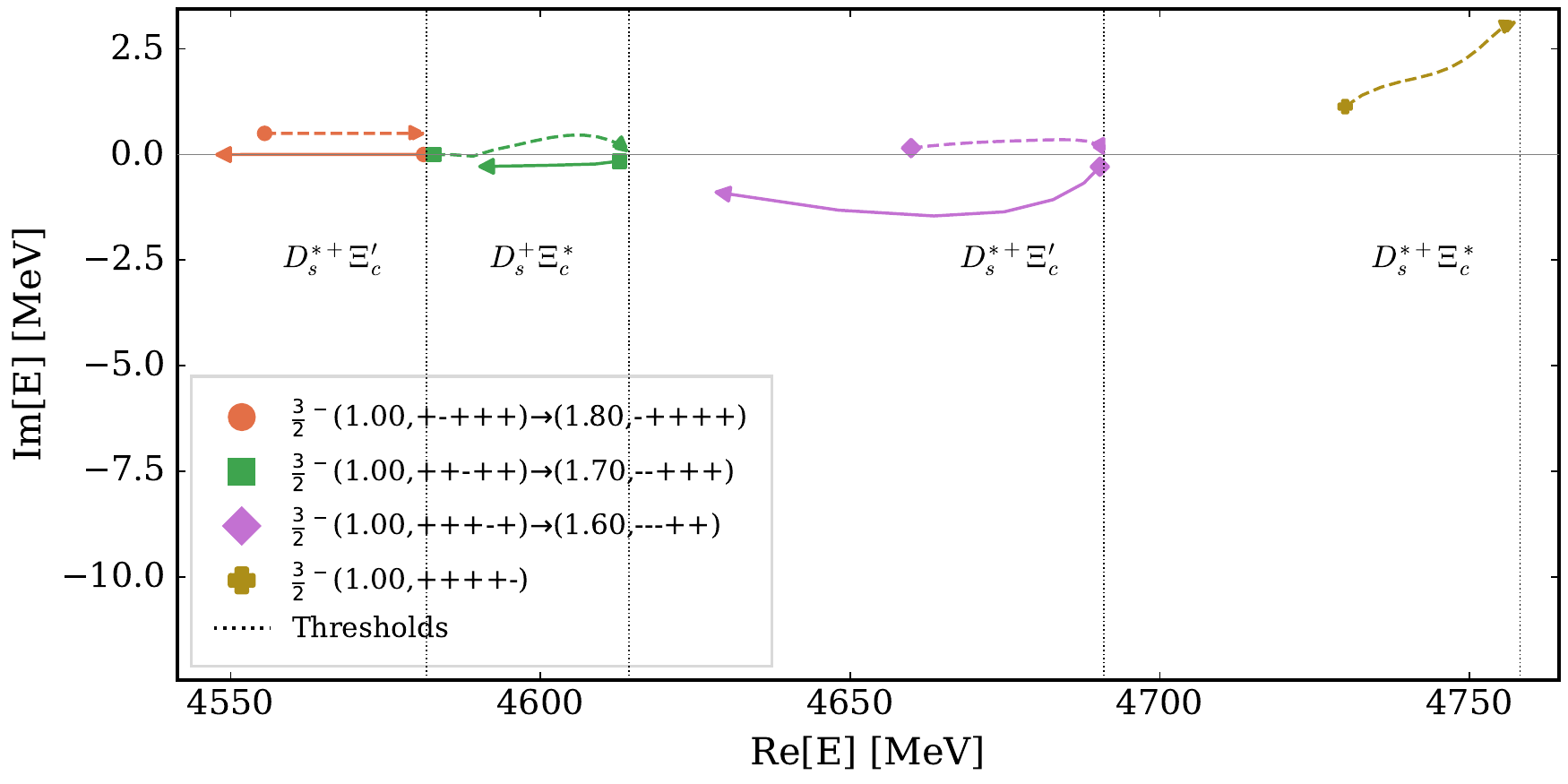}
	\caption{Trajectory of the near-threshold poles in $D_s^{*+}\Xi_c$-$D_s^{+}\Xi_c^*$-$D_s^{*+}\Xi_c^{'}$-$D_s^{*+}\Xi_c^{*}$ channels with $J^P=3/2^-$ by varying the cutoff. The trajectory of the virtual pole of $D_s^{*+}\Xi_c$ system is artificially moved from the real axis to the complex plane for better illustration. See the caption of Fig.~\ref{fig:RSS10}.}\label{fig:RSS30}
\end{figure}

\begin{figure}[ht]\centering
	\includegraphics[width=0.48\textwidth]{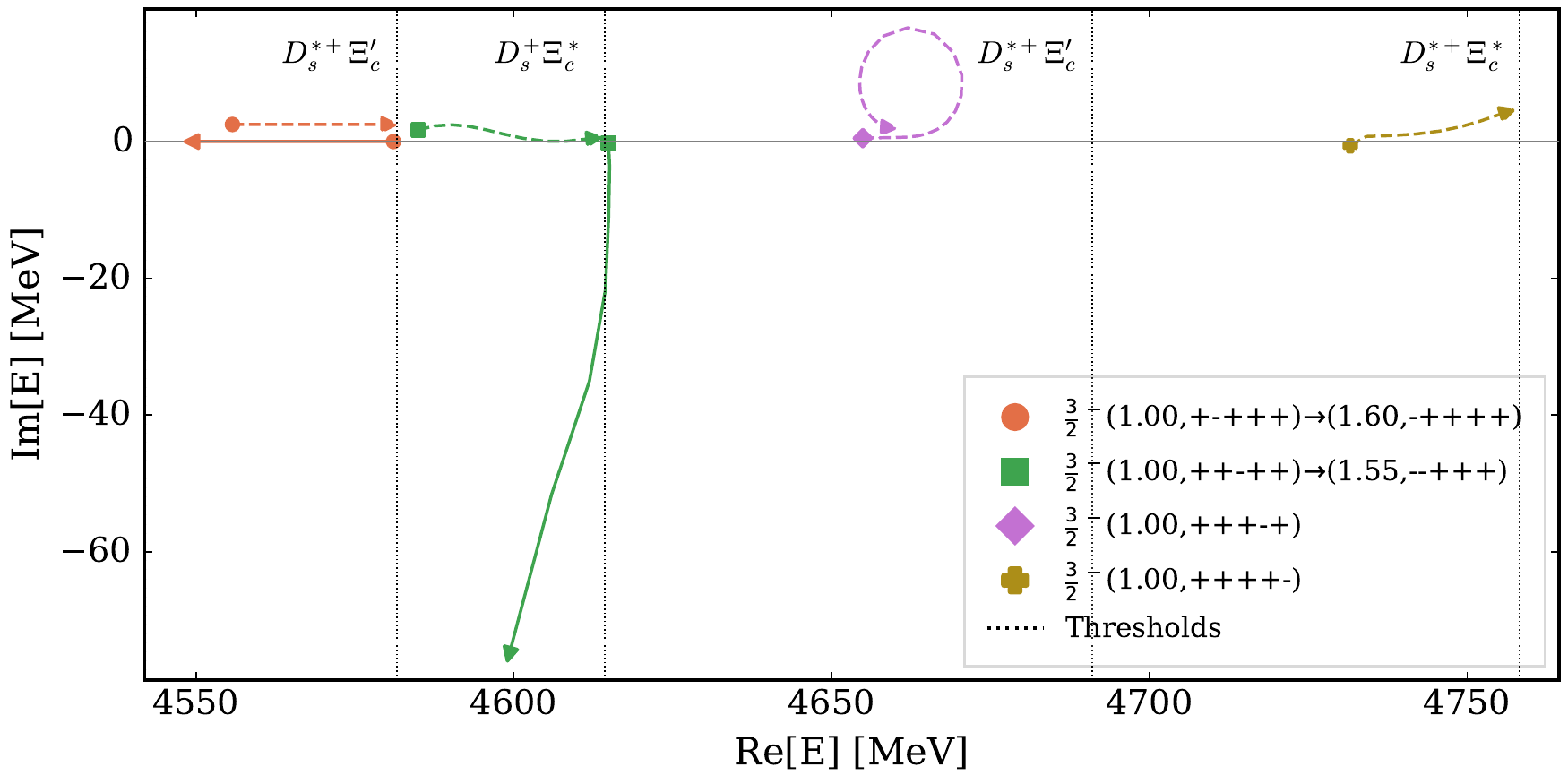}
	\caption{Similar figure as Fig.~\ref{fig:RSS30} after removing the $\delta(\bm r)$-term. }\label{fig:RSS31}
\end{figure}

For the $J^P=3/2^-$ system, four channels, $D_s^{*+}\Xi_c$-$D_s^{+}\Xi_c^*$-$D_s^{*+}\Xi_c^{'}$-$D_s^{*+}\Xi_c^{*}$, can couple in $S$-wave, and the pole trajectories near the thresholds of these four channels as the cutoff increases from $1.0$ GeV are shown in Fig.~\ref{fig:RSS30}, and the pole trajectories after removing the $\delta(\bm r)$-term are shown in Fig.~\ref{fig:RSS31}. In both cases of including and excluding $\delta(\bm r)$-terms, the pole below the threshold of $D_s^{*+}\Xi_c^{*}$ channel would not appear on the RS connected to the physical real axis within the cutoff range $1.0-3.0$ GeV.
\footnote{ We have noticed the strange behavior of the pole close to and below the threshold of $D_s^{*+}\Xi_c^{'}$ when $a=1$ in Fig.~\ref{fig:RSS31} and we have verified it by varying $a$ from 0.5 to 1.0 in a step of 0.1. }

Since only $D_s^{*+}\Xi_c^{*}$ can form the $J^P=5/2^-$ system if only $S$-wave is considered, no coupled-channel dynamics appears. 

To estimate the contribution of the possible $D$-wave components, we turn on the $S$-$D$-wave mixing potential and calculate pole positions near the thresholds of $D_s^{+}\Xi_c$, $D_s^{+}\Xi_c^{'}$, $D_s^{*+}\Xi_c$, $D_s^{+}\Xi_c^{*}$, $D_s^{*+}\Xi_c^{'}$ and $D_s^{*+}\Xi_c^{*}$ channels. The behaviors of these poles by varying the cutoff are presented in Appendix~\ref{app:SD-wave}, which indicates that $S$-$D$-wave mixing effects make the poles to appear on the RS connected to the physical real energy axis with smaller cutoff. In another word, $S$-$D$-wave mixing provides additional attractions to these systems. Especially, such effects are more important for the poles with $J^P=3/2^-,5/2^-$ near thresholds of $D_s^{*+}\Xi_c^{'}$ and $D_s^{*+}\Xi_c^{*}$ channels. 

Basically, in our calculation, we can determine neither the cutoff $\Lambda$ nor the reduction parameter $a$ which represents the contribution of the $\delta(\bm r)$-term, i.e., the short-range interaction, because there is not any experimental data for double-charm pentaquarks with hidden strangeness. However, as an illustrative result, we could present the full coupled-channel results including $S$-$D$-wave mixing effects here by fixing the cutoff to $1.5$ GeV, which is somehow phenomenologically reasonable as the LHCb $P_c$ pentaquarks~\cite{LHCb:2019kea} are reproduced with $\Lambda=1.4$ GeV in Ref.~\cite{Yalikun:2021bfm}, with $\Lambda=1.04$ and 1.32 GeV in Ref.~\cite{Chen:2019asm}. In Ref.~\cite{Tornqvist:1993ng}, it is mentioned that in nucleon-nucleon interactions the values of $\Lambda$ between 0.8 and 1.5 GeV have been used depending on the model and application, and the larger values ($\Lambda > 1.5$ GeV) are also required for nucleon-nucleon phase shifts. By setting $\Lambda=1.5$ GeV, ten poles are found near the thresholds of $D_s^{+}\Xi_c$, $D_s^{+}\Xi_c^{'}$, $D_s^{*+}\Xi_c$, $D_s^{+}\Xi_c^{*}$, $D_s^{*+}\Xi_c^{'}$ and $D_s^{*+}\Xi_c^{*}$ channels, as shown in Table~\ref{tab:RSSD1500a0}. Among them, the pole below the threshold of the lowest channel $D_s^{+}\Xi_c$ with $J^P=1/2^-$ is a bound state. The poles with imaginary parts except for the ones labeled with superscripts ``$\bigtriangleup$" are resonances. These poles correspond to the solid lines in Figs.~\ref{fig:RSS10} and \ref{fig:RSS30}, which are directly connected to the physical real axis and therefore, called by ``bound-state-like" poles or physical resonances. On the contrary, the poles with superscripts ``$\bigtriangleup$", corresponding to the dashed lines in Figs.~\ref{fig:RSS10} and \ref{fig:RSS30}, are not directly connected to the physical real axis and therefore, called by ``virtual-state-like" poles. We want to emphasize that the  ``virtual-state-like"  pole, although located on the RS remote from the physical real axis, may cause clear cusp (peak- or dip-like) structure at the threshold if the pole is located close to the threshold. The resonances can decay to lower channels and the partial decay widths shown in the last column of Table~\ref{tab:RSSD1500a0} are calculated using the procedure presented in Ref.~\cite{Yalikun:2021dpk}. Similar results obtained by removing the $\delta(\bm r)$-term from the potentials are shown in Table~\ref{tab:RSSD1500a1}. For the cases without the $\delta(\bm r)$, compared to the cases with $\delta(\bm r)$-term, it is found that 1) $1/2^-(D_s^{*+}\Xi_c^{'})$, $1/2^-(D_s^{*+}\Xi_c^{*})$ and $3/2^-(D_s^{*+}\Xi_c^{*})$ become resonances; 2) $3/2^-(D_s^{+}\Xi_c^{*})$, $3/2^-(D_s^{*+}\Xi_c^{*})$ and $5/2^-(D_s^{*+}\Xi_c^{*})$ turn to ``virtual-state-like" poles. 

 From the former results, it can be seen that some states, such as $1/2^-(D_s^{+}\Xi_c^{'})$ and $3/2^-(D_s^{+}\Xi_c^{*})$, extremely close to their thresholds, and they will appear as resonances if the additional attraction due to the exchange of $\eta'$ and $f_0(980)$ mesons. Consequently, the pole positions for other states also slightly move away from the corresponding thresholds due to such attraction. Up to the masses of exchanged mesons, $\eta'$ and $f_0(980)$ exchanges in the $D_s^{(*)+}\Xi_c^{(*,')}$ systems are the same as $\eta$ and $\sigma$ exchanges, respectively, and the potentials of these systems are described in the appendix~\ref{sec:potential}. For their coupling constants, we assume that the $f_0(980)$ coupling has the same strength with $\sigma$ coupling, and the universal couplings of the pseudoscalar octet mesons $g_1$ and $g$ are adopted for the couplings of $\eta'$ by including it into $\rm{SU(3)}$ octet of pseudoscalar meson~\cite{Aceti:2014uea}. Masses of $\eta'$ and $f_0(980)$ are taken as $957.8$ and $990.0$ MeV\cite{Zyla:2020zbs}. With cutoff $\Lambda=1.5~\rm{GeV}$, pole positions for the $1/2^-(D_s^{+}\Xi_c)$, $1/2^-(D_s^{+}\Xi_c^{'})$, $1/2^-(D_s^{*+}\Xi_c)$, $3/2^-(D_s^{*+}\Xi_c)$, $3/2^-(D_s^{+}\Xi_c^{*})$, $3/2^-(D_s^{*+}\Xi_c^{'})$, $5/2^-(D_s^{*+}\Xi_c^{*})$ states in this case are $4436.81$, $4546.45-i0.03$, $4558.57-i1.38$, $4564.9-i0.11$, $4612.86-i0.3$, $4684.02-i3.6$ and $4716.58-i12.22$ in the units of $\rm{MeV}$, respectively. Compared to the results in Table~\ref{tab:RSSD1500a0}, the pole corresponding to $1/2^-(D_s^{+}\Xi_c^{'})$ state appears at the RS connected to the physical real energy axis, and other poles are shifted towards the lower thresholds. In the case without $\delta(r)$-term, the poles labeled with $1/2^-(D_s^{+}\Xi_c^{'})$ and $3/2^-(D_s^{+}\Xi_c^{*})$ appear in the RS connected to the physical real energy axis at the positions of $(4544.84 - i0.22)\rm{ ~MeV}$ and $(4613.99 - i0.44)\rm{~MeV}$, respectively. Other poles are also slightly shifted towards the lower threshold.

Note that in the above calculations, $D_s^{+}\Xi_c$ is the lowest channel and therefore, the pole below its threshold is located on the real axis and is stable against strong interaction. Actually, $D_s^{+}\Xi_c$, as well as other higher channels, can transit into $D^{(*)}\Lambda_c$ or $D^{(*)}\Sigma_c^{(*)}$ via $K^{(*)}$ exchange, which will lead to a finite width of the $D_s^{+}\Xi_c$ bound state. Since we are not aiming at a precise result, we introduce only $D^{(*)}\Lambda_c$ channels into the previous coupled channels to roughly estimate such decay width of $1/2^-(D_s^{+}\Xi_c)$ bound state.\footnote{Compared to $D^{(*)}\Sigma_c^{(*)}$, $D^{(*)}\Lambda_c$ has a larger phase space and it is also easier to be detected in experiments.} The potentials for $D^{(*)}\Lambda_c$ channels coupled to $D_s^{(*)+}\Xi_c^{(*,')}$ channels are listed in appendix~\ref{app:other-pot}. The $1/2^-(D_s^{+}\Xi_c)$ bound state now moves into the complex energy plane and the pole position together with its partial decay widths is shown in Table~\ref{tab:decay-BS}. We can find that the sum of partial decay widths of $1/2^-(D_s^{+}\Xi_c)$ bound state to $D^{(*)}\Lambda_c$ final states is ${\cal O}(1\ \rm MeV)$ . For the poles related to other higher channels, we expected similar or larger contributions of $D^{(*)}\Lambda_c$ channels to their decay widths due to the larger phase space. Therefore, we consider $D^{(*)}\Lambda_c$ as good places to search for the predicted states in the  $D_s^{+}\Xi_c$-$D_s^{+}\Xi_c^{'}$-$D_s^{*+}\Xi_c$-$D_s^{+}\Xi_c^{*}$-$D_s^{*+}\Xi_c^{'}$-$D_s^{*+}\Xi_c^{*}$ coupled-channel system. 
  
\begin{table*}[ht]\centering
    \caption{Pole positions and partial decay widths of the states in the coupled-channel system of $D_s^{+}\Xi_c$-$D_s^{+}\Xi_c^{'}$-$D_s^{*+}\Xi_c$-$D_s^{+}\Xi_c^{*}$-$D_s^{*+}\Xi_c^{'}$-$D_s^{*+}\Xi_c^{*}$ when $\Lambda=1.5$ GeV with $S$-$D$-wave mixing. The poles labeled with the superscript ``$\bigtriangleup$" are the ``virtual-state-like" poles emerged on the RSs far away from the physical real axis. Each entry labeled with ``$\cdots$" in the column $\Gamma_i$ means that the decay is not allowed. }\label{tab:RSSD1500a0}
    \begin{ruledtabular}\centering
    \begin{tabular}{ccccc}
        $J^P$ & Nearby channel & Threshold [MeV] & $E_{\rm pole}$ [MeV] & $\Gamma_i$($D_s^+\Xi_c/D_s^+\Xi_c'/D_s^{*+}\Xi_c/D_s^+\Xi_c^*/D_s^{*+}\Xi_c/D_s^{*+}\Xi_c^*$) [MeV]\\ \hline
		\multirow{5}{*}{$1/2^-$}
		& $D_s^+\Xi_c$ & 4437.76&$4437.71$ & $\cdots$\\
		& $D_s^{+}\Xi_c'$& 4547.14& $4547.04-i0.01^\bigtriangleup$ & $\cdots$ \\
		& $D_s^{*+}\Xi_c$ &4581.62 &$4564.26-i1.00$ & $0.18/1.81/\cdots/\cdots/\cdots/\cdots$ \\
		& $D_s^{*+}\Xi_c'$ & 4691.00&$4687.07-i3.97^\bigtriangleup$ & $\cdots$ \\
		& $D_s^{*+}\Xi_c^*$ &4758.17 &$4754.05-i4.27^\bigtriangleup$ & $\cdots$ \\ \hline
		\multirow{4}{*}{$3/2^-$}
		& $D_s^{*+}\Xi_c$ &4581.62 &$4569.56-i0.02$ & $0.01/0.04/\cdots/\cdots/\cdots/\cdots$ \\
		& $D_s^+\Xi_c^*$ &4614.31& $4614.29-i0.05$ & $0.00/0.02/0.10/\cdots/\cdots/\cdots$ \\
		& $D_s^{*+}\Xi_c'$ &4691.00& $4689.01-i2.58$ & $3.36/0.06/1.9/0.36/\cdots/\cdots$ \\
		& $D_s^{*+}\Xi_c^*$&4758.17 & $4769.34-i9.95^\bigtriangleup$ & $\cdots$ \\ \hline
		$5/2^-$& $D_s^{*+}\Xi_c^*$&4758.17 & $4727.40-i13.37$ & $7.82/0.19/19.27/0.33/0.02/\cdots$
	\end{tabular}
    \end{ruledtabular}
\end{table*}
 
\begin{table*}[ht]\centering
    \caption{Same as Table.~\ref{tab:RSSD1500a0} but without $\delta(\bm r)$-term.}\label{tab:RSSD1500a1}
    \begin{ruledtabular}\centering
	\begin{tabular}{ccccc}
            $J^P$ & Nearby channel & Threshold [MeV] & $E_{\rm pole}$ [MeV] & $\Gamma_i$($D_s^+\Xi_c/D_s^+\Xi_c'/D_s^{*+}\Xi_c/D_s^+\Xi_c^*/D_s^{*+}\Xi_c/D_s^{*+}\Xi_c^*$) [MeV]\\ \hline
			\multirow{5}{*}{$1/2^-$}
			& $D_s^+\Xi_c$& 4437.76 & $4437.73$ & $\cdots$\\
			& $D_s^{+}\Xi_c'$& 4547.14 & $4547.14-i0.00^\bigtriangleup$ & $\cdots$ \\
			& $D_s^{*+}\Xi_c$ &4581.62 & $4565.34-i2.68$ & $0.18/4.98/\cdots/\cdots/\cdots/\cdots$ \\
			& $D_s^{*+}\Xi_c'$& 4691.00 & $4686.30-i4.49$ & $1.20/6.41/2.02/0.01/\cdots/\cdots$ \\
			& $D_s^{*+}\Xi_c^*$&4758.17 & $4742.51-i6.44$ & $2.81/2.57/6.26/0.05/1.46/\cdots$ \\ \hline
			\multirow{4}{*}{$3/2^-$}
			& $D_s^{*+}\Xi_c$ &4581.62 & $4570.09-i0.02$ & $0.00/0.04/\cdots/\cdots/\cdots/\cdots$ \\
			& $D_s^+\Xi_c^*$&4614.31 & $4614.26-i0.22^\bigtriangleup$ & $\cdots$ \\
			& $D_s^{*+}\Xi_c'$&4691.00 & $4689.71-i6.38^\bigtriangleup$ & $\cdots$ \\
			& $D_s^{*+}\Xi_c^*$&4758.17 & $4747.06-i16.76$ & $2.29/0.02/24.51/7.32/3.89/\cdots$ \\ \hline
			$5/2^-$& $D_s^{*+}\Xi_c^*$&4758.17 & $4763.39-i11.20^\bigtriangleup$ & $\cdots$
	\end{tabular}
    \end{ruledtabular}
\end{table*}

\begin{table}[ht]\centering
    \caption{Pole positions and partial decay widths of the $1/2^-(D_s^{+}\Xi_c)$ bound state after including the lower $D^{(*)}\Lambda_c$ channels. The $\delta({\bm r})$-term and the $S$-$D$ mixing are considered.}\label{tab:decay-BS}
    \begin{ruledtabular}\centering
    \begin{tabular}{ccc}
    $\Lambda$[MeV]&$E_{\rm{pole}}$[MeV]&$\Gamma_i(D\Lambda_c/D^*\Lambda_c)$[MeV]\\
      $1500$ &  $4437.62-i0.37$ & $0.88/0.02$\\
   $1600$ &  $4435.92 - i1.12$ & $2.04/0.06$\\
    \end{tabular}
    \end{ruledtabular}
\end{table}
\section{Summary}\label{sec:summary}


In this work, as partners of the $P_c$ pentaquarks, the double-charm hidden-strangeness pentaquarks near the $D_s^{+(*)}\Xi_c^{(',*)} $ thresholds are systematically investigated in the hadronic molecular picture. Possible near-threshold states as their molecular candidates are explored within the OBE model. First, the possible bound or virtual states in six single channels $D_s^{+}\Xi_c$, $D_s^{+}\Xi_c^{'}$, $D_s^{*+}\Xi_c$, $D_s^{+}\Xi_c^{*}$, $D_s^{*+}\Xi_c^{'}$ and $D_s^{*+}\Xi_c^{*}$ are calculated by solving the Schr\"odinger equation with OBE potentials, including the $S$-$D$-wave mixing. By varying the cutoff in the range of $1.0-2.5$ GeV and including the $\delta(\bm r)$-term, five states with $J^P=1/2^-$, four states with $J^P=3/2^-$ and one state with $J^P=5/2^-$ can form virtual states if $\Lambda>1$ GeV and they turn to bound states when $\Lambda$ gets large enough.   In addition, the results after removing the $\delta(\bm r)$-term are also presented. Second, the coupled-channel dynamics of $D_s^{+}\Xi_c$-$D_s^{+}\Xi_c^{'}$-$D_s^{*+}\Xi_c$-$D_s^{+}\Xi_c^{*}$-$D_s^{*+}\Xi_c^{'}$-$D_s^{*+}\Xi_c^{*}$ is further investigated, and masses and widths of ten possible resonances and bound states as the molecular candidates for double-charm hidden-strangeness pentaquarks are calculated. For the ten molecular states in our coupled-channel analysis, the role of $\delta(\bm r)$-term in the OBE potentials is also examined. Its influence on the poles near thresholds of $D_s^{*+}\Xi_c^{'}$ and $D_s^{*+}\Xi_c^{*}$ channels is more significant. Our work indicates that among these ten poles, the pole with $J^P=1/2^-$ below $D_s^{+}\Xi_c$ threshold is a bound state, which becomes a resonance after introducing the coupling to lower $D^{(*)}\Lambda_c$ and $D^{(*)}\Sigma_c^{(*)}$ channels, two poles with $J^P=1/2^-$ and $3/2^-$ below the threshold of $D_s^{*+}\Xi_c$ channel are physical resonances and other seven poles are resonances or ``virtual-state-like" poles, depending on the contribution of the $\delta(\bm r)$-term in the OBE model. Further experimental investigations are required to verify these results. These poles could lead to near-threshold structures in the $D^{(*)}\Lambda_c$ final states and can be searched for in the future.

\begin{acknowledgments}
 
 This work is supported by the Doctoral Program of Tian Chi Foundation of Xinjiang Uyghur Autonomous Region of China under grant No. 51052300506, by the NSFC and the Deutsche Forschungsgemeinschaftn(DFG, German Research Foundation) through the funds provided to the Sino-German Collaborative Research Center TRR110 “Symmetries and the Emergence of Structure in QCD” (NSFC Grant No. 12070131001, DFG Project-ID 196253076 - TRR 110), by the NSFC Grant No.11835015, No.12047503, and by the Chinese Academy of Sciences (CAS) under
Grant No.XDB34030000.

\end{acknowledgments}

 \section*{Appendix}
 \begin{appendices}
 \section{Potentials related to $D_s^{(*)}\Xi_c^{(',*)}$ channels}\label{sec:potential}
 We collect the potentials related to $D_s^{(*)}\Xi_c^{(',*)}$ channels in the following.
 \begin{subequations}\label{eq:poten-in-p}
 \begin{align}
\mathcal{V}^{11}&=2l_Bg_S \frac{\chi_3^\dagger  \chi_1 }{\bm q^2+m_\sigma^2}-\frac{\beta\beta_Bg_V^2 }{2} \frac{\chi_3^\dagger  \chi_1}{\bm q^2+m_\phi^2},\\
\mathcal{V}^{15}&=\frac{gg_4}{\sqrt{6}f_\pi^2}\frac{(\chi^\dagger_3 \bm\sigma\chi_1\cdot \bm q) (\bm\epsilon_4^* \cdot\bm q)}{\bm q^2+\mu^2_{\eta}}\notag\\
&+\frac{2\lambda\lambda_Ig_V^2}{\sqrt{6}}\frac{(\chi_3^\dagger\bm\sigma\chi_1\times\bm q)\cdot (\bm\epsilon_4^*\times\bm q)}{\bm q^2+\mu^2_{\phi}},\label{eq:p-v15}\\
\mathcal{V}^{16}&=-\frac{gg_4}{\sqrt{2}f_\pi^2}\frac{(\bm\chi^\dagger_3\cdot \bm q
\chi_1)(\bm\epsilon_{4}^*\cdot\bm q)}{\bm q^2+\mu^2_{\eta}}\notag\\
&-\sqrt{2}\lambda\lambda_Ig_V^2 \frac{(\bm\chi_3^{\dagger} \times \bm q \chi_1)\cdot(\bm \epsilon_4^{*}\times \bm q)}{\bm q^2+\mu^2_{\phi}},\\
\mathcal{V}^{22}&=-l_Sg_S\frac{\chi_3^\dagger\chi_1}{\bm q^2+m_\sigma^2}+\frac{\beta\beta_Sg_V^2}{4}\frac{\chi^\dagger_3\chi_1}{\bm q^2+m_\phi^2},\\
\mathcal{V}^{23}&=\frac{gg_4}{\sqrt{6}f_\pi^2}\frac{(\chi^\dagger_3 \bm \sigma\chi_1\cdot \bm q) (\bm\epsilon_2 \cdot\bm q)}{\bm q^2+\mu^2_{\eta}}\notag\\
&+\frac{2\lambda\lambda_Ig_V^2}{\sqrt{6}}\frac{(\chi_3^\dagger\bm\sigma\chi_1\times\bm q)\cdot (\bm\epsilon_2\times\bm q)}{\bm q^2+\mu^2_{\phi}},\\
\mathcal{V}^{24}&=\frac{l_Sg_S}{\sqrt 3}\frac{\bm \chi^{\dagger}_3\cdot\bm\sigma\chi_1}{\bm q^2+\mu_{\sigma}^2}-\frac{\beta\beta_Sg_V^2}{2\sqrt 3}\frac{\bm\chi_3^\dagger\cdot\bm\sigma\chi_1}{\bm q^2+\mu^2_{\phi}},\\
\mathcal{V}^{25}&=\frac{gg_1}{6f_\pi^2}\frac{(\chi_3^\dagger\bm\sigma\chi_1\cdot\bm q)(\bm\epsilon_4^*\cdot\bm q)}{\bm q^2+\mu_{\eta}^2}\notag\\
&+\frac{\lambda\lambda_S g_V^2}{3}\frac{(\chi_3^\dagger\bm\sigma\chi_1\times \bm q)\cdot (\bm \epsilon_4^{*}\times \bm q)}{\bm q^2+\mu_{\phi}^2},\\
\mathcal{V}^{26}&=\frac{\sqrt 3 g g_1}{12f_\pi^2}\frac{(i\bm\chi^\dagger_{3} \times\bm\sigma\chi_1)\cdot\bm q\bm\epsilon_4^{*}\cdot \bm q}{\bm q^2+\mu_{\eta}^2}\notag\\
&+\frac{\lambda\lambda_Sg_V^2}{2\sqrt 3}\frac{(i\bm\chi^\dagger_3\times\bm\sigma\chi_1\times \bm q)\cdot(\bm\epsilon_4^{*}\times \bm q)}{\bm q^2+\mu_{\phi}^2},\\
\mathcal{V}^{33}&=2l_Bg_S \frac{\chi_3^\dagger  \chi_1 \bm\epsilon_4^* \cdot\bm\epsilon_2}{\bm q^2+m_\sigma^2}-\frac{\beta\beta_Bg_V^2}{2} \frac{\chi_3^\dagger  \chi_1 \bm\epsilon_4^* \cdot\bm\epsilon_2}{\bm q^2+m_\phi^2},\\
\mathcal{V}^{34}&=-\frac{gg_4}{\sqrt{2}f_\pi^2}\frac{(\bm\chi^\dagger_3\cdot \bm q
\chi_1)(\bm\epsilon_{2}\cdot\bm q)}{\bm q^2+\mu^2_{\eta}}\notag\\
&-\frac{2\lambda\lambda_Ig_V^2}{\sqrt 2} \frac{(\bm\chi_3^{\dagger} \times \bm q \chi_1)\cdot(\bm \epsilon_2\times \bm q)}{\bm q^2+\mu^2_{\phi}},\\
\mathcal{V}^{35}&=\frac{gg_4}{\sqrt{6}f_\pi^2}\frac{(\chi^\dagger_3\bm\sigma\chi_1\cdot \bm q) (i\bm \epsilon_{2}\times\bm\epsilon_{4}^*)\cdot \bm q}{\bm q^2+\mu_{\eta}^2}\notag\\
&+\frac{2\lambda_I\lambda g_V^2}{\sqrt 6}\frac{(\chi_3^\dagger\bm\sigma\chi_1\times \bm q)\cdot(i\bm\epsilon_{2}\times \bm\epsilon_{4}^{*}\times\bm q)}{\bm q^2+\mu_{\phi}^2},\\
\mathcal{V}^{36}&=-\frac{gg_4}{\sqrt 2}\frac{\bm \chi_3^\dagger \cdot \bm q\chi_1(i\bm \epsilon_{2}\times\bm \epsilon_{4}^*)\cdot \bm q}{\bm q^2+\mu^2_{\eta}}\notag\\
&-\sqrt 2\lambda\lambda_I g_V^2\frac{(\bm \chi^{\dagger}_3 \times\bm q \chi_1)\cdot (i\bm\epsilon_{2}\times\bm\epsilon_{4}^{*}\cdot \bm q)}{\bm q^2+\mu_{\phi}^2},\\
\mathcal{V}^{44}&=-l_Sg_S\frac{\bm\chi^\dagger_3\cdot\bm\chi_1}{\bm q^2+m_\sigma^2}+\frac{\beta\beta_S g_V^2}{4}\frac{\bm\chi^\dagger_3\cdot\bm\chi_1}{\bm q^2+m_\phi^2},\\
\mathcal{V}^{45}&=\frac{g g_1}{4\sqrt 3f_\pi^2}\frac{(i\chi^\dagger_3\bm\sigma\times \bm \chi_{1})\cdot \bm q(\bm \epsilon_{4}^*\cdot\bm q )}{\bm q^2+\mu_{\eta}^2}\notag\\
&+\frac{\lambda\lambda_Sg_V^2}{2\sqrt 3}\frac{(i\chi_3^\dagger\bm\sigma\times \bm\chi_1\times\bm q )\cdot (\bm\epsilon_{4}^{*}\times \bm q)}{\bm q^2+\mu_{\phi}^2},\\
\mathcal{V}^{46}&=-\frac{g g_1}{4f_\pi^2}\frac{(i\bm\chi^\dagger_3\times\bm\chi_1)\cdot \bm q \bm\epsilon_{4}^*\cdot\bm q}{\bm q^2+\mu_{\eta}^2}\notag\\
&-\frac{\lambda\lambda_Sg_V^2}{2}\frac{(i\bm\chi^\dagger_3\times\bm\chi_1\times\bm q)\cdot(\bm\epsilon_{4}^*\times\bm q)}{\bm q^2+\mu_{\phi}^2},\\
\mathcal{V}^{55}&=-l_Sg_S\frac{\chi_3^\dagger\chi_1\bm\epsilon_{4}^*\cdot\bm\epsilon_{2}}{\bm q^2+m_\sigma^2}+\frac{\beta\beta_Sg_V^2}{4}\frac{\chi_3^\dagger\chi_1(\bm\epsilon_{4}^{*}\cdot\bm\epsilon_{2})}{\bm q^2+m_\phi^2}\notag\\
&+\frac{gg_1}{6f_\pi}\frac{\chi^\dagger_3\bm\sigma\chi_1\cdot\bm q(i\bm\epsilon_{2}\times\bm\epsilon_{4}^*)\cdot \bm q}{\bm q^2+m_\eta^2}\notag\\
&+\frac{\lambda\lambda_Sg_V^2}{3}\frac{(\chi_3^\dagger\bm\sigma\chi_1\times \bm q)\cdot(i\bm\epsilon_{2}\times\bm\epsilon_{4}^*\times \bm q)}{\bm q^2+m_\phi^2},\\
\mathcal{V}^{56}&=\frac{ g g_1}{4\sqrt 3f_\pi^2}\frac{(i\bm\chi^\dagger_{3}\times\bm\sigma\chi_1)\cdot\bm q(i\bm\epsilon_{2}\times\bm\epsilon_{4}^*)\cdot\bm q}{\bm q^2+\mu_{\eta}^2}\notag\\
&+\frac{\lambda\lambda_Sg_V^2}{2\sqrt 3}\frac{(i\bm\chi_3^\dagger\times \bm\sigma\chi_1\times\bm q)\cdot(i\bm\epsilon_{2}\times\bm\epsilon_{4}^{*} \times\bm q )}{\bm q^2+\mu_{\phi}^2},\\
\mathcal{V}^{66}&=-l_Sg_S\frac{(\bm\chi^\dagger_3\cdot\bm\chi_1)(\bm\epsilon_{4}^{*}\cdot\bm\epsilon_{2})}{\bm q^2+m_\sigma^2}+\frac{\beta\beta_Sg_V^2}{4}\frac{(\bm\chi_{3}^\dagger \cdot \bm\chi_1)(\bm\epsilon_{4}^{*}\cdot\bm\epsilon_{2})}{\bm q^2+m_\phi^2}\notag\\
&-\frac{g g_1}{4f_\pi^2}\frac{(i\bm\chi_{3}^\dagger\times\bm\chi_{1})\cdot \bm q(i\bm\epsilon_{2}\times\bm\epsilon_{4}^*)\cdot\bm q}{\bm q^2+m_\eta^2}\notag\\
&-\frac{\lambda\lambda_Sg_V^2}{2}\frac{(i\bm\chi_3^\dagger\times\bm \chi_1\times \bm q)\cdot(i\bm \epsilon_{2}\times\bm\epsilon_{4}^{*}\times\bm q)}{\bm q^2+m_\phi^2},
 \end{align}
\end{subequations}
 where $\bm \epsilon_2$ and $\bm \epsilon_4^*$ are the polarization vectors for charmed mesons in the initial and final states. 

 \section{Variations of pole positions with $S$-$D$-wave mixing}\label{app:SD-wave}
In this section, we show the pole behaviours with $S$-$D$-wave mixing when varying $\Lambda$. In the coupled-channel system with $J^P=1/2^-$, five poles are found as the cutoff varies from $1.45$ to $2.6$ GeV, and their positions are given in Table ~\ref{tab:RSP12}, where the sign of the imaginary part of each channel momentum is shown in the parenthesis. The pole labeled with $E^{I}_{\rm{pole}}$ is located at the real energy axis on RS-I, and it is a bound state. The poles labeled with $E^{\rm{II}}_{\rm{pole}}$ and $E^{\rm{III}}_{\rm{pole}}$ together with $E^{I}_{\rm{pole}}$ emerge with relatively smaller cutoff compared to other two poles labeled with $E^{\rm{VI}}_{\rm{pole}}$ and $E^{\rm{V}}_{\rm{pole}}$, and the latter two are much more broad.  

 \begin{table*}[ht]\centering
 \caption{Pole positions on the RSs close to the physical real axis in the coupled-channel system with $J^P=1/2^-$. Each entry with a ``$\cdots$'' means that the pole goes to other RS far away from the physical real axis. $\Lambda$ and pole position($E_{\rm{pole}}^{\rm{RS}}$) are in unit of MeV.}\label{tab:RSP12}
 \begin{ruledtabular}\centering
 \begin{tabular}{cccc|ccc}
 $\Lambda$&$E_{\rm{pole}}^{\rm{I}}(++++++)$&$E_{\rm{pole}}^{\rm{II}}(-+++++)$&$E_{\rm{pole}}^{\rm{III}}(--++++)$&$\Lambda$&$E_{\rm{pole}}^{\rm{V}}(----++)$&$E_{\rm{pole}}^{\rm{VI}}(-----+)$\\
 \hline
 $1450.0$ & $\cdots $ & $ \cdots        $ & $4572.86-i0.69$ & $ 2450.0 $ & $ 4690.37-i2.08  $ & $ 4736.88-i26.43$ \\ 
 $1500.0$ & $4437.71$ & $ \cdots        $ & $4564.26-i1.00$ & $ 2500.0 $ & $ 4688.48-i4.84  $ & $ 4731.39-i35.46$ \\ 
 $1550.0$ & $4436.76$ & $ 4546.13-i0.05 $ & $4552.59-i1.62$ & $ 2550.0 $ & $ 4684.87-i12.06 $ & $ 4726.27-i46.64$ \\
 $1600.0$ & $4434.43$ & $ 4545.90-i0.01 $ & $4536.67-i0.27$ & $ 2600.0 $ & $ 4675.30-i24.38 $ & $ 4722.37-i59.79$ \\
 \end{tabular}
 \end{ruledtabular}
 \end{table*}
 
 If the $\delta(\bm r)$-term in the potentials is removed, the positions of the five poles in $J^P=1/2^-$ system as the cutoff increases are shown in Table ~\ref{tab:RSP121}. In this sector, considerable effects of the $\delta(\bm r)$-term can be seen from the last three poles labeled as $E_{\rm{pole}}^{\rm{III}}$, $E_{\rm{pole}}^{\rm{V}}$ and $E_{\rm{pole}}^{\rm{VI}}$. For instance, after removing the $\delta(\bm r)$-term, the third pole labeled with $E_{\rm{pole}}^{\rm{III}}$ becomes broad, the fourth and fifth poles labeled with $E_{\rm{pole}}^{\rm{V}}$ and $E_{\rm{pole}}^{\rm{VI}}$ emerges in the region of RSs connected to the physical real energy axis with relatively smaller cutoff, because these poles are relevant to the bound states in the single channel analysis shown in Fig.~\ref{fig:BS-VS}.   
 
 \begin{table*}[ht]\centering
 \caption{Same as Table~\ref{tab:RSP12} but the $\delta(\bm r)$-term is removed.}\label{tab:RSP121}
 \begin{ruledtabular}\centering
 \begin{tabular}{cccccc}
 $\Lambda$&$E_{\rm{pole}}^{\rm{I}}(++++++)$&$E_{\rm{pole}}^{\rm{II}}(-+++++)$&$E_{\rm{pole}}^{\rm{III}}(--++++)$&$E_{\rm{pole}}^{\rm{V}}(----++)$&$E_{\rm{pole}}^{\rm{VI}}(-----+)$\\
 \hline
 $1400.0$ &  $\cdots $ &  $\cdots        $ & $4579.28-i0.09$ &  $4690.86-i0.79 $ &  $4754.72-i2.14$\\
 $1500.0$ &  $4437.73$ &  $\cdots        $ & $4565.34-i2.68$ &  $4686.30-i4.49 $ &  $4742.51-i6.44$\\
 $1600.0$ &  $4433.68$ &  $\cdots        $ & $4546.50-i7.00$ &  $4676.58-i10.59$ &  $4722.74-i12.84$\\
 $1700.0$ &  $4419.55$ &  $ 4545.88-i0.01$ & $4490.75-i0.74$ &  $4661.41-i18.69$ &  $4697.24-i20.24$\\
 \end{tabular}
 \end{ruledtabular}
 \end{table*}
 
 For $J^P=3/2$ system, as shown in Table ~\ref{tab:RSP32}, four poles are found as the cutoff increases from $1.4$ to $1.55$ GeV. Three of them labeled with $E_{\rm{pole}}^{\rm{III}}$, $E_{\rm{pole}}^{\rm{IV}}$ and $E_{\rm{pole}}^{\rm{V}}$ are below the thresholds of $D_s^{*+}\Xi_c$, $D_s^{+}\Xi_c^{*}$ and $D_s^{*+}\Xi_c^{'}$ channels, while the last one labeled with $E_{\rm{pole}}^{\rm{VI}}$ is above the threshold of  $D_s^{+}\Xi_c^{*}$ channel. After removing the $\delta(\bm r)$-term, the pole $E_{\rm{pole}}^{\rm{III}}$ keeps almost the same, and the cutoff dependence of the other poles in this sector is also changed as it is shown in Fig.~\ref{tab:RSP321}.  
 
 For the $J^P=5/2^-$ system, only one pole is found. The positions of the pole considering the coupled-channel potential with or without the $\delta(\bm r)$-term are shown in Table ~\ref{tab:RSP52}. Compared to the poles near the thresholds of $D_s^{*+}\Xi_c'$ and $D_s^{*+}\Xi_c^*$ channels in coupled-channel systems with $J^P=1/2^-$ and $3/2^-$, the effect of the $\delta(\bm r)$-term on the pole in $J^P=5/2^-$ sector is not much significant. Among these ten poles, $E_{\rm{pole}}^{\rm{I,II,III}}$ poles in $J^P=1/2^-$ system, $E_{\rm{pole}}^{\rm{III,IV}}$ poles in $J^P=3/2^-$ system and $E_{\rm{pole}}^{\rm{VI}}$ pole in $J^P=5/2^-$ system are not sensitive to the $\delta(\bm r)$-term, and they can appear with relativity smaller cutoff, while the other poles near thresholds of $D_s^{*+}\Xi_c'$ and $D_s^{*+}\Xi_c^*$ channels in $J^P=1/2^-,3/2^-$ systems have different behaviors in the case of with or without the $\delta(\bm r)$-term. In each case, the poles near thresholds of $D_s^{*+}\Xi_c'$ and $D_s^{*+}\Xi_c^*$ channels are broad and difficult to be observed.
 
 \begin{table*}[ht]\centering
 \caption{Same as Table~\ref{tab:RSP12} but with $J^P=3/2^-$. }\label{tab:RSP32}
 \begin{ruledtabular}\centering
 \begin{tabular}{ccccc}
 $\Lambda$&$E_{\rm{pole}}^{\rm{III}}$(--++++)&$E_{\rm{pole}}^{\rm{IV}}$(---+++)&$E_{\rm{pole}}^{\rm{V}}$(----++)&$E_{\rm{pole}}^{\rm{VI}}$(-----+)\\
 \hline
 $1400.0$ & $4579.95-i0.02$ & $ \cdots        $ & $\cdots       $ &  $ \cdots        $ \\ 
 $1450.0$ & $4576.07-i0.03$ & $ \cdots        $ & $4691.46-i0.79$ &  $ 4767.11-i2.60 $ \\ 
 $1500.0$ & $4569.56-i0.02$ & $ 4614.29-i0.05 $ & $4689.01-i2.58$ &  $ 4769.34-i9.95 $ \\
 $1550.0$ & $4560.07-i0.01$ & $ 4612.57-i0.28 $ & $4683.28-i4.09$ &  $ 4771.33-i20.23$ \\
 \end{tabular}
 \end{ruledtabular}
 \end{table*}
 
 \begin{table*}[ht]\centering
 \caption{Same as Table~\ref{tab:RSP32} but the $\delta(\bm r)$-term is removed.}\label{tab:RSP321}
 \begin{ruledtabular}\centering
 \begin{tabular}{ccccc}
 $\Lambda$&$E_{\rm{pole}}^{\rm{III}}$(--++++)&$E_{\rm{pole}}^{\rm{IV}}$(---+++)&$E_{\rm{pole}}^{\rm{V}}$(----++)&$E_{\rm{pole}}^{\rm{VI}}$(-----+)\\
 \hline
 $1500.0$ &  $4570.09-i0.02$ & $ \cdots      $ & $ \cdots      $ & $4747.06-i16.76$\\
 $1600.0$ &  $4543.47-i0.03$ & $4613.45-i1.14$ & $ \cdots      $ & $4723.38-i25.74$\\
 $1700.0$ &  $4511.62-i0.00$ & $4606.96-i3.86$ & $ \cdots      $ & $4693.00-i27.76$\\
 $1780.0$ &  $4509.34-i0.00$ & $4594.41-i5.62$ & $4697.89-i1.80$ & $4666.32-i23.34$\\
 \end{tabular}
 \end{ruledtabular}
 \end{table*}
 \begin{table}[H]\centering
 \caption{Position of a pole in the coupled-channel system with $J^P=5/2^-$. $\Lambda$ and pole position($E_{\rm{pole}}^{\rm{RS}}$) are in unit of MeV.}\label{tab:RSP52}
 \begin{ruledtabular}\centering
 \begin{tabular}{cc|cc}
	 \multicolumn{2}{c}{With $\delta(\bm r)$} & \multicolumn{2}{c}{Without $\delta(\bm r)$}\\ \hline
 $\Lambda$&$E_{\rm{pole}}^{\rm{VI}}(-----+)$&$\Lambda$&$E_{\rm{pole}}^{\rm{VI}}(-----+)$\\
 \hline
 $1350.0$ & $4759.44-i4.93 $ & $1400.0$ & $4767.33-i0.00$\\
 $1400.0$ & $4753.08-i9.21 $ & $1500.0$ & $4763.39-i11.20$\\
 $1450.0$ & $4742.35-i12.21$ & $1600.0$ & $4745.55-i21.61$\\
 $1500.0$ & $4727.40-i13.37$ & $1700.0$ & $4717.43-i23.05$\\
 \end{tabular}
 \end{ruledtabular}
 \end{table}
 
 \section{Potentials related to $D^{(*)}\Lambda_c$ channels}\label{app:other-pot}
 With the procedure described in Subsec.~\ref{subsec_Lag}, the effective potentials in the momentum space for $D^{(*)}\Lambda_c$ channels coupled to $D_s^{(*)+}\Xi_c^{(',*)}$ channels are derived using the Lagrangian in Eq.~\eqref{lag}, and shown in the following. 
 \begin{align}
\mathcal{V}^{D\Lambda_c\to D\Lambda_c}&=2l_Bg_S \frac{\chi_3^\dagger  \chi_1 }{\bm q^2+m_\sigma^2}-\frac{\beta\beta_Bg_V^2 }{2} \frac{\chi_3^\dagger  \chi_1}{\bm q^2+m_\omega^2},\\
\mathcal{V}^{D\Lambda_c\to D_s^+\Xi_c}&=-\frac{\beta\beta_Bg_V^2 }{2} \frac{\chi_3^\dagger  \chi_1}{\bm q^2+\mu^2_{K^*}},\\
\mathcal{V}^{D\Lambda_c\to D_s^{*+}\Xi_c'}&=\frac{gg_4}{\sqrt{6}f_\pi^2}\frac{(\chi^\dagger_3 \bm\sigma\chi_1\cdot \bm q) (\bm\epsilon_4^* \cdot\bm q)}{\bm q^2+\mu^2_{K}}\notag\\
&+\frac{2\lambda\lambda_Ig_V^2}{\sqrt{6}}\frac{(\chi_3^\dagger\bm\sigma\chi_1\times\bm q)\cdot (\bm\epsilon_4^*\times\bm q)}{\bm q^2+\mu^2_{K^*}},\\
\mathcal{V}^{D\Lambda_c\to D_s^{*+}\Xi_c^*}&=-\frac{gg_4}{\sqrt{2}f_\pi^2}\frac{(\bm\chi^\dagger_3\cdot \bm q
\chi_1)(\bm\epsilon_{4}^*\cdot\bm q)}{\bm q^2+\mu^2_{K}}\notag\\
&-\sqrt{2}\lambda\lambda_Ig_V^2 \frac{(\bm\chi_3^{\dagger} \times \bm q \chi_1)\cdot(\bm \epsilon_4^{*}\times \bm q)}{\bm q^2+\mu^2_{K^*}},\\
\mathcal{V}^{D^*\Lambda_c\to D^*\Lambda_c}&=2l_Bg_S \frac{\chi_3^\dagger  \chi_1 \bm\epsilon_4^* \cdot\bm\epsilon_2}{\bm q^2+m_\sigma^2}-\frac{\beta\beta_Bg_V^2}{2} \frac{\chi_3^\dagger  \chi_1 \bm\epsilon_4^* \cdot\bm\epsilon_2}{\bm q^2+m_\omega^2},\\
\mathcal{V}^{D^*\Lambda_c\to D^{*+}_s\Xi_c}&=-\frac{\beta\beta_Bg_V^2}{2} \frac{\chi_3^\dagger  \chi_1 \bm\epsilon_4^* \cdot\bm\epsilon_2}{\bm q^2+\mu_{K^*}},\\
\mathcal{V}^{D^*\Lambda_c\to D^{+}_s\Xi_c'}&=\frac{gg_4}{\sqrt{6}f_\pi^2}\frac{(\chi^\dagger_3 \bm \sigma\chi_1\cdot \bm q) (\bm\epsilon_2 \cdot\bm q)}{\bm q^2+\mu^2_{K}}\notag\\
&+\frac{2\lambda\lambda_Ig_V^2}{\sqrt{6}}\frac{(\chi_3^\dagger\bm\sigma\chi_1\times\bm q)\cdot (\bm\epsilon_2\times\bm q)}{\bm q^2+\mu^2_{K^*}},\\
\mathcal{V}^{D^*\Lambda_c\to D^{+}_s\Xi_c^*}&=-\frac{gg_4}{\sqrt{2}f_\pi^2}\frac{(\bm\chi^\dagger_3\cdot \bm q
\chi_1)(\bm\epsilon_{2}\cdot\bm q)}{\bm q^2+\mu^2_{K}}\notag\\
&-\frac{2\lambda\lambda_Ig_V^2}{\sqrt 2} \frac{(\bm\chi_3^{\dagger} \times \bm q \chi_1)\cdot(\bm \epsilon_2\times \bm q)}{\bm q^2+\mu^2_{K^*}},\\
\mathcal{V}^{D^*\Lambda_c\to D^{*+}_s\Xi_c'}&=\frac{gg_4}{\sqrt{6}f_\pi^2}\frac{(\chi^\dagger_3\bm\sigma\chi_1\cdot \bm q) (i\bm \epsilon_{2}\times\bm\epsilon_{4}^*)\cdot \bm q}{\bm q^2+\mu_{K}^2}\notag\\
&+\frac{2\lambda_I\lambda g_V^2}{\sqrt 6}\frac{(\chi_3^\dagger\bm\sigma\chi_1\times \bm q)\cdot(i\bm\epsilon_{2}\times \bm\epsilon_{4}^{*}\times\bm q)}{\bm q^2+\mu_{K^*}^2},\\
\mathcal{V}^{D^*\Lambda_c\to D^{*+}_s\Xi_c^*}&=-\frac{gg_4}{\sqrt 2}\frac{\bm \chi_3^\dagger \cdot \bm q\chi_1(i\bm \epsilon_{2}\times\bm \epsilon_{4}^*)\cdot \bm q}{\bm q^2+\mu^2_{K}}\notag\\
&-\sqrt 2\lambda\lambda_I g_V^2\frac{(\bm \chi^{\dagger}_3 \times\bm q \chi_1)\cdot (i\bm\epsilon_{2}\times\bm\epsilon_{4}^{*}\cdot \bm q)}{\bm q^2+\mu_{K^*}^2},\\
 \end{align}
 where the masses of ralevent particles are $m_\omega=782.7$ MeV, $m_K=493.7$ MeV and $m_{K^*}=891.7$ MeV.

 \end{appendices}
 
\bibliographystyle{apsrev4-1}
\bibliography{main.bib}
\end{document}